\documentclass[english]{cupconf}
\usepackage[T1]{fontenc}
\usepackage[latin1]{inputenc}
\usepackage{varioref}
\usepackage{color}
\usepackage{graphicx}
\usepackage{amssymb}
\usepackage[authoryear]{natbib}

\makeatletter

\providecommand{\tabularnewline}{\\}

\usepackage{times}
 
  \IfFileExists{amsbsy.sty}
    {\typeout{^^JFound the 'amsbsy' package on the system, using it.^^J}%
     \usepackage{amsbsy}}
    {}

\newsavebox{\astrutbox}
\sbox{\astrutbox}{\rule[-5pt]{0pt}{20pt}}

\usepackage{jour-abbrev}

\usepackage{babel}
\makeatother

\begin{document}
\newcommand{\Mo}{M_{\odot}}
\newcommand{\Ro}{R_{\odot}}
\newcommand{\Lo}{L_{\odot}}
\newcommand{\Mbh}{M_{\bullet}}
\newcommand{\Ms}{M_{\star}}
\newcommand{\Rs}{R_{\star}}
\newcommand{\Ls}{L_{\star}}
\newcommand{\Ns}{N_{\star}}
\newcommand{\nstr}{n_{\star}}
\newcommand{\Np}{N_{p}}
\newcommand{\Mp}{M_{p}}
\newcommand{\Rp}{R_{p}}

\title[Relaxation Processes Near the Galactic MBH]{Stellar Relaxation
Processes Near the Galactic Massive Black Hole\footnote{Invited
talk. To appear in ``2007 STScI spring symposium: Black Holes'', eds,
M. Livio \& A. M. Koekemoer, Cambridge University Press, in press.}}

\author[T. Alexander]{Tal Alexander$^{1,2}$}
\affiliation{$^{1}$Faculty of Physics, Weizmann Institute of Science, PO box 26, Rehovot 76100, Israel\\
$^{2}$William Z. and Eda Bess Novick Career Development Chair}
\maketitle

\begin{abstract}
The massive black hole (MBH) in the Galactic Center and the stars
around it form a unique stellar dynamics laboratory for studying how
relaxation processes affect the distribution of stars and compact
remnants and lead to close interactions between them and the MBH.
Recent theoretical studies suggest that processes beyond {}``minimal''
two-body relaxation may operate and even dominate relaxation and its
consequences in the Galactic Center. I describe loss-cone refilling
by massive perturbers, strong mass segregation and resonant relaxation;
review observational evidence that these processes play a role in
the Galactic Center; and discuss some cosmic implications for the
rates of gravitational wave emission events from compact remnants
inspiraling into MBHs, and the coalescence timescales of binary MBHs. 
\end{abstract}

\section{Introduction}

The $\Mbh\!\sim\!4\times10^{6}\,\Mo$ massive black hole (MBH) in
the Galactic Center (GC) \citep{eis+05,ghe+05} and the stars around
it are the closest and observationally most accessible of such systems.
Observations of the GC thus offer a unique opportunity to study in
great detail the effects of the MBH and its extreme environment on
star formation, stellar evolution and stellar dynamics, and the interactions
of stars and compact remnants with the MBH. 

Here the focus is stellar relaxation processes. Relaxation plays an
important role in a wide range of phenomena that involve close interactions
with a MBH (the {}``loss-cone problem'', \S\ref{ss:LC}). Such
processes include gravitational wave (GW) emission by compact remnants
inspiraling into a MBH ({}``extreme-mass ratio inspiral events''
(EMRIs, see review by \citealt{ama+07}, tidal flares from tidal
disruption events \citep{fra+76}, tidal capture and tidal scattering
of stars \citep{ale+03a,ale+01b}, 3-body exchanges with binaries
leading to the capture of stars on tight orbits around the MBH and
the ejection of hyper velocity stars (HVSs) out of the galaxy \citep{hil88},
the orbital decay and coalescence of a binary MBHs (and the {}``last
parsec stalling problem'', see review by \citealt{mer+05b}), and
perhaps also the formation of ultra-luminous X-ray sources (ULXs)
in star clusters following stellar capture around an intermediate
mass black hole (IMBH) \citep{hop+04}. Relaxation processes are possibly
linked to the presence and properties of unusual stellar populations
that are observed near MBHs, such as the central {}``S-star'' cluster,
the stellar disks in the GC \citep{eis+05,pau+06} and the stellar
disk in M31 \citep{ben+05}. 

Dynamical relaxation by star-star interactions is inherent to the
discreteness of stellar systems. In the absence of additional mechanisms
to randomize stars in phase-space, standard 2-body stellar relaxation
assures a minimal degree of randomization, albeit one that could be
too slow to be of practical interest. This review will discuss relaxation
processes beyond standard stellar relaxation, which operate on much
shorter timescales, or else operate in a qualitatively different way:
massive perturbers (\S \ref{s:MP}), strong mass segregation (\S
\ref{s:MSeg}) and resonant relaxation (\S \ref{s:RR}). 

The dynamical properties of the GC, specifically its short 2-body
relaxation time and high stellar density, are probably not typical
of galaxies in general (\S\ref{ss:GCdyn}). However, dynamical processes
that can be probed by GC observations have implications beyond the
GC. In particular, the Milky Way is the archetype of the subset of
galaxies with low-mass MBHs that are key targets for planned space-borne
gravitational wave detectors, such as the Laser Interferometer Space
Antenna (LISA). GC studies may help understand the effect of such
relaxation processes on the open questions of the cosmic EMRI event
rate and the EMRI orbital characteristics.

Before turning to a discussion of the non-standard relaxation processes
that are expected to operate in the GC, it is useful to briefly review
the dynamics leading to close interactions with a MBH (loss-cone theory)
and the dynamical conditions in the GC.

\subsection{Infall and inspiral into a MBH}

\label{ss:LC}

Stars can fall into the MBH either by losing orbital energy, so that
the orbit shrinks down to the size of the last stable circular orbit
($r_{\mathrm{LSCO}}\!=\!3r_{s}$ for a non-rotating MBH, where the
event horizon is at the Schwarzschild radius $r_{s}\!=\!2G\Mbh/c^{2}$
), or by losing orbital angular momentum so that the orbit becomes
nearly radial and unstable (periapse $r_{p}\!<\!2r_{s}$ for a star
with zero orbital energy falling into a non-rotating MBH)%
\footnote{If the stars are tidally disrupted before falling in the MBH, the
relevant distance scale is the tidal disruption radius $r_{t}\!\sim\!\Rs(\Mbh/\Ms)^{1/3}\!>\! r_{s}$
rather than the event horizon $r_{s}$.%
}. The timescale to lose energy by 2-body scattering, $T_{E}\!\equiv\!|E/\dot{E}|$
is of the order of the relaxation time, \begin{equation}
T_{E}\sim T_{R}\sim\left(M_{\bullet}/M_{\star}\right)^{2}\tau_{\mathrm{dyn}}(r)/N_{\star}(<r)\log N_{\star}(<r)\,,\label{e:TE}\end{equation}
where $\Ns(<r)$ is the number of stars inside $r$, $\tau_{\mathrm{dyn}}(r)\!\sim\!\sqrt{r^{3}/G\Mbh}$
is the dynamical time and spherical symmetry and a Keplerian velocity
dispersion are assumed, $\sigma^{2}\!\sim\! G\Mbh/r$. The maximal
angular momentum available for an orbit with energy $E$ is that of
a circular orbit, $J_{c}(E)\!=\! G\Mbh/\sqrt{2E}$ (using here the
stellar dynamical sign convention $E\!\equiv\!-v^{2}/2-\phi(r)\!>\!0$).
The timescale for losing angular momentum, $T_{J}\!\equiv\!|J/\dot{J}|$,
can be much shorter than $T_{E}$ when $J\!<\! J_{c}$, since

\textcolor{black}{\begin{equation}
T_{J}=[J/J_{c}(E)]^{2}T_{E}\,.\label{e:TJ}\end{equation}
As a consequence, almost all stars that reach the MBH, and are ultimately
destroyed by a close interaction with it, do so by being scattered
to low-$J$ {}``loss-cone'' orbits (near radial orbits with $J\!<\! J_{lc}\!\simeq\!\sqrt{2G\Mbh q}$,
where $q$ is the maximal periapse required for the close interaction
of interest to occur. \citealt{fra+76,lig+77}). The rate of close
interaction events, $\Gamma_{\mathrm{lc}}$, is set by the replenishment
rate of stars into the loss-cone. When the replenishment mechanism
is diffusion in phase space by 2-body scattering, $\Gamma_{\mathrm{lc}}\!\propto\! T_{R}^{-1},$
which is typically a very low rate. Close to the MBH, at high-$E$,
where the relative size of the loss-cone in phase-space is large ($J_{lc}/J_{c}\!\propto\!\sqrt{qE}$),
relaxation is too slow to replenish the lost stars, and the loss-cone
is on average empty. Farther out, at low $E$, where the loss-cone
is small, relaxation can replenish the lost stars, the loss-cone is
full (isotropic distribution of stars) and the local replenishment
rate is maximal. Nevertheless, the contribution to the total replenishment
rate from the low-$E$, full loss-cone regions of phase-space, where
the timescales are longer and the stellar densities lower, remains
small compared to that from the empty loss-cone regions at high-$E$
\citep{lig+77}. }

\textcolor{black}{The observational and theoretical interest in such
close interactions motivated numerous investigations of alternative
efficient loss-cone replenishment mechanism, such as 2-body relaxation
in non-spherically symmetric potentials \citep{mag+99,ber+06}, chaotic
orbits in triaxial potentials \citep{nor+83,mer+04b,ger+85,hol+06},
relaxation by massive perturbers \citep{zha+02,per+07}, resonant
relaxation \citep{rau+96, rau+98, hop+06a, lev07}, or perturbations
by a massive accretion disk or a secondary IMBH \citep{pol+94, lev+05}. }

Close interactions with a MBH fall in two dynamical categories \citep{ale+03b}:
infall processes, such as tidal disruption, where the star is destroyed
promptly on its first close encounter with the MBH, and inspiral processes,
such as GW EMRI events, where multiple consecutive close encounters
are required for the orbit to gradually decay. The infall takes about
an orbital period, the time to fall from the point of deflection to
the center, whereas the inspiral process takes much longer, depending
on the energy extraction efficiency of the dissipational mechanism
involved (for example GW emission, tidal heating or drag against a
massive accretion disk). In most cases the dissipated energy is a
steeply decreasing function of the the periapse%
\footnote{E.g. the GW energy emitted per orbit scales as $\Delta E\!\propto\!(\Ms c^{2}/\Mbh)(r_{p}/r_{s})^{-7/2}$
\citep{pet64}.%
} and so the inspiral time scales with the number of periapse passages,
and hence with the initial orbital period. 

\begin{figure}
\noindent \begin{raggedright}
\begin{tabular}{cc}
\includegraphics[width=0.5\columnwidth,keepaspectratio]{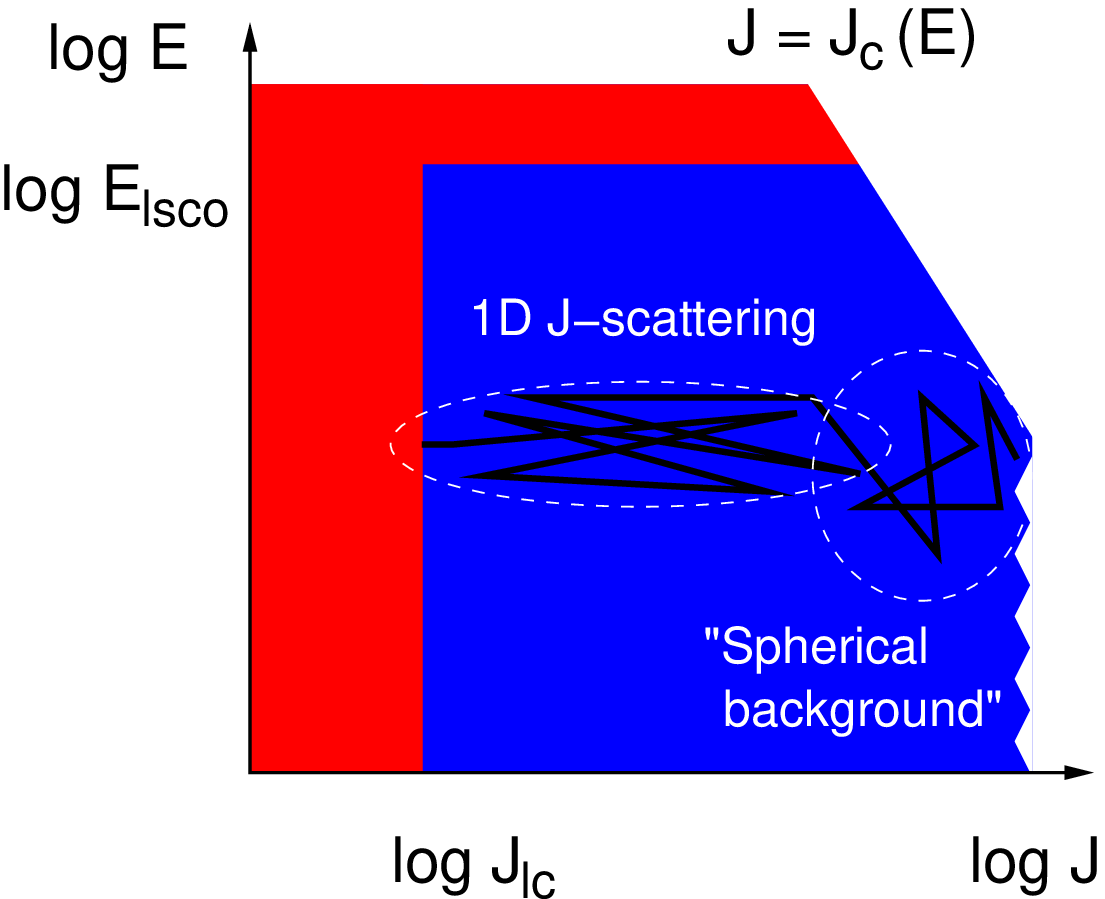} & \includegraphics[width=0.5\columnwidth,keepaspectratio]{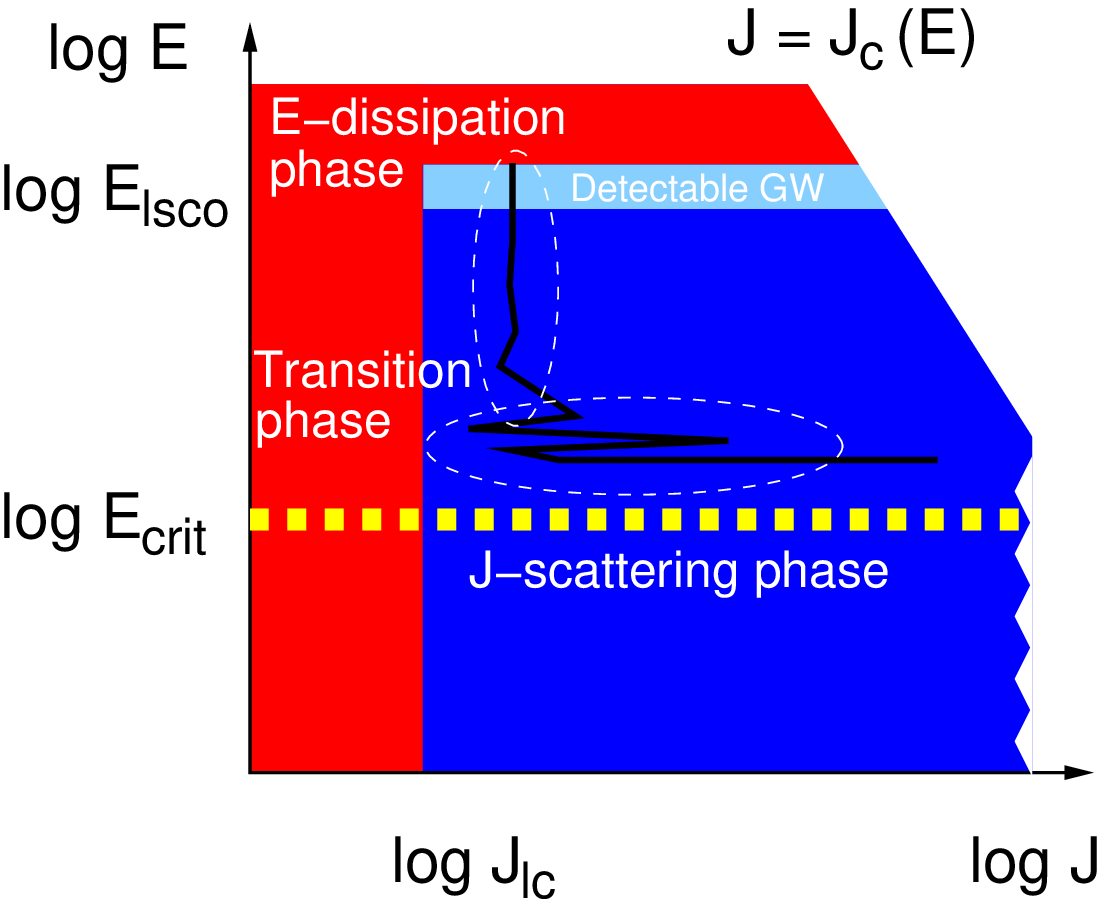}\tabularnewline
\end{tabular}
\par\end{raggedright}

\caption{\label{f:rcrit} A schematic representation of the phase-space $(\log E,\log J)$
trajectories leading a star to the MBH. Each segment of the random-walk
trajectory represents the change in the phase coordinates over some
fixed time step $\Delta t$. The shaded areas on top ($E\!>\! E_{\mathrm{LSCO}}$)
and on the left ($J\!<\! J_{\mathrm{lc}}$) are regions of phase space
where no stable orbits exist. The diagonal boundary on the right is
s the maximal angular momentum $J_{c}(E)$. Left: Infall without dissipation.
A star with initially high $J$ is scattered with roughly equal relative
magnitude in $E$ and $J$. Eventually a random kick will send it
to a low-$J$ orbit, where $J$-scattering is much faster than $E$-scattering,
making it plunge directly into the MBH. Right: Inspiral with dissipation.
Energy dissipation by the emission of GW can lead to very rapid orbital
decay on low-$J$ orbits, faster than the mean time between scattering
events, thus enabling the star to reach a short-period orbit with
detectable GW emission (narrow horizontal shaded strip on top). Statistically,
nearly all stars with initial energy $E\!>\! E_{\mathrm{crit}}$ will
ultimately inspiral into the MBH, while nearly all stars with $E\!<\! E_{\mathrm{crit}}$
will ultimately plunge into the MBH, following a trajectory similar
to the one depicted in the left panel. }

\end{figure}

An infall or inspiral event can occur only if the star, once deflected
into the loss-cone, avoids being re-scattered out of it (and in the
case of inspiral, also avoids being scattered directly into the MBH).
Because inspiral processes are slow, stars can avoid re-scattering,
complete the inspiral and decay to an interesting, very short period
orbit with high emitted dissipative power, only if they are deflected
into the loss cone from an initially short period orbit, with $E\!>\! E_{\mathrm{crit}}$.
Figure (\ref{f:rcrit}) shows a schematic description of the phase-space
evolution of infalling and inspiraling stars, and the emergence of
a critical energy scale. For inspiral by GW emission into a $\Mbh\!\sim\! O(10^{6}\,\Mo)$
MBH, $E_{\mathrm{crit}}$ corresponds to an initial distance scale
of $r_{\mathrm{crit}}\sim\!0.01$ pc (the \emph{ansatz} $r\!\leftrightarrow\! E\!=\! G\Mbh/2a$\textcolor{black}{,}
is assumed here\textcolor{black}{, where} $a$ \textcolor{black}{is
the Keplerian semi-major axis}). The EMRI event rate is then approximately
\citep{hop+05} \begin{equation}
\Gamma_{\mathrm{lc}}\sim N_{\mathrm{GW}}(<r_{\mathrm{crit}})/T_{R}(r_{\mathrm{crit}})\propto N_{\mathrm{GW}}(<r_{\mathrm{crit}})\Ns(<r_{\mathrm{crit}})/\tau_{\mathrm{dyn}}(r_{\mathrm{crit}})\,,\label{e:Glc}\end{equation}
where $N_{\mathrm{GW}}(<r)$ is the number of potential GW sources
(compact remnants) within distance $r$ of the MBH. A critical energy
can be similarly defined for infall processes. Because infall is much
faster than inspiral, $E_{\mathrm{crit}}$ is much lower ($r_{\mathrm{crit}}$
much larger). For example, the critical radius for tidal disruption
in the GC is $r_{\mathrm{crit}}\!\sim\mathrm{few\,}\mathrm{pc}$ \citep{lig+77,sye+99,mag+99}.
Most of the stars that infall or inspiral originate near $r_{\mathrm{crit}}$.

Equation (\ref{e:Glc}) shows that the degree of central concentration
of compact remnants strongly affects the EMRI event rate. Mass segregation
therefore substantially increases the predicted EMRI event rate from
inspiraling ${\cal O}(10\,\Mo)$ stellar black holes (SBHs), which
are the most massive, long-lived objects in the population (\citealt{hop+06b};
\S \ref{s:MSeg}). Similarly, the capture of compact remnants very
near the MBH by 3-body exchanges between the MBH and binaries (\S
\ref{s:MP}) can also strongly affect the EMRI rate (Perets, Hopman
\& Alexander, 2007, in prep.). The dependence of $\Gamma_{\mathrm{lc}}$
on $T_{R}$ is not trivial, since $r_{\mathrm{crit}}$ itself depends
on $T_{R}$: the shorter the relaxation time, the faster stars are
scattered into the loss-cone, but also out of it. Detailed analysis
shows that the two effects cancel out for $n_{\star}\!\propto\! r^{-3/2}$
stellar cusps. Since in most galactic nuclei the logarithmic slope
of the density profile is not much different from $-3/2$, the EMRI
rate is expected to be roughly independent of the relaxation time
\citep{hop+05}. It should be emphasized that this result applies
only to 2-body relaxation, and needs to be re-examined if other loss-cone
replenishment mechanisms dominate the dynamics.

\subsection{The dynamical state of the stellar system around the Galactic MBH}

\label{ss:GCdyn}

The stellar system around the Galactic MBH is expected to be in a
state of dynamical relaxation in a high density cusp. This is a direct
consequence of the low mass of the Galactic MBH and of the $\Mbh/\sigma$
relation, the tight observed correlation between the mass of central
MBHs and the typical velocity dispersion in the bulges of their host
galaxies, $\Mbh\!\propto\!\sigma^{\beta}$, where $4\!\lesssim\!\beta\!\lesssim\!5$
\citep{fer+00,geb+00}. $\beta\!=\!4$ is assumed here for simplicity;
the conclusions below are reinforced if $\beta\!>\!4$. 

The MBH radius of dynamical influence is conventionally defined as
$r_{h}\!\sim\! GM_{\bullet}/\sigma^{2}\!\propto\!\Mbh^{1/2}$. The
mass in stars within the radius of influence is of the order of the
mass of the MBH, so their number is $N_{h}\!\sim\! M_{\bullet}/M_{\star}$,
where $\Ms$ is the mean stellar mass, and the average stellar density
within $r_{h}$ is $\bar{n}_{h}\!\sim\! N_{h}/r_{h}^{3}$. The two-body
relaxation time at $r_{h}$ is $T_{R}\!\sim\!\left(M_{\bullet}/M_{\star}\right)^{2}\tau_{h}/N_{h}$.
It then follows that $T_{R}\!\propto\!\Mbh^{5/4}$ and $\bar{n}_{h}\!\propto\! M_{\bullet}^{-1/2}$.
Evaluated for the Galactic MBH, $T_{R}\!\sim\!{\cal O}(1\,\mathrm{Gyr})\!<t_{H}$
(the Hubble time) and $\bar{n}_{h}\!\sim\!{\cal O}(10^{5}\,\mathrm{pc^{-3}})$.
The short relaxation time implies that the system will return to its
relaxed steady state following a major perturbation, such as a merger
with a second MBH \citep{mer+06,mer+07}. Note that $T_{R}\!>\! t_{h}$
for a MBH only a few times more massive than the Galactic MBH. The
GC is thus a member of a relatively small subset of galaxies with
high-density relaxed stellar cusps.

A relaxed stellar system is expected to settle into a power-law cusp
distribution, $n_{\star}\!\propto\! r^{-\alpha}$, (\S \ref{s:MSeg}).
The high stellar density in a steeply rising cusp allows star-star
and star-MBH interactions to become frequent enough to be dynamically
relevant and observationally interesting (Eq. \ref{e:Glc}). For example,
the rates of both tidal disruption events \citep{wan+04} and EMRI
inspiral events \citep{hop+05} scale inversely with the MBH mass,
$\Gamma\!\propto\! N_{h}/T_{R}\!\propto\!\Mbh^{-1/4}$.

\section{Massive perturbers}

\label{s:MP}

\subsection{Massive perturbers and the loss-cone}

\label{ss:MP_LC}

The relaxation time (Eq. \ref{e:TE}) is proportional to $(\Ms^{2}\nstr)^{-1}$.
This can be readily understood by considering the {}``$\Gamma\!\sim\! nv\Sigma$''
collision rate between stars of mass $\Ms$ and mean space density
in volume V, $\nstr\!=\!\Ns/V$, where the cross-section $\Sigma\sim\pi r_{c}^{2}$
is evaluated for collisions at the capture radius $r_{c}\!=\!2G\Ms/v^{2}$,
the minimal radius for a soft encounter with a typical velocity $v$.
The rate of scattering by stars is then $\Gamma_{\star}\!\sim\nstr\Ms^{2}/v^{3}\!\sim\! T_{R}^{-1}$
(integration over all collision radii increases the rate only by a
logarithmic Coulomb factor). When the system also contains a few very
massive objects such as giant molecular clouds (GMCs), stellar clusters,
or IMBHs (if such exist), these massive perturbers (MPs) of mass $M_{p}\!\gg\!\Ms$
and space density $n_{p}\!=\! N_{p}/V\ll\!\nstr$ will scatter stars
at the capture radius $r_{c}\!=\! G(\Ms+M_{p})/v^{2}$ at a rate of
$\Gamma_{p}\!\sim\! n_{p}(\Ms+M_{p})^{2}/v^{3}\!\sim n_{p}M_{p}^{2}/v^{3}$.
MPs could well dominate the relaxation even if they are very rare,
as long as \begin{equation}
\mu_{2}\!\equiv\! M_{p}^{2}N_{p}/\Ms^{2}\Ns\!>\!1\,.\label{e:mu2MP}\end{equation}

Efficient relaxation by MPs was first suggested by \citet{spi+51,spi+53}
to explain stellar velocities in the Galactic disk. Its relevance
for replenishing the loss-cone was subsequently investigated in the
context of Solar system dynamics for the scattering of Oort cloud
comets to the Sun \citep{hil81,bai83}, and more recently as a mechanism
for establishing the $\Mbh/\sigma$ correlation by fast accretion
of stars and dark matter \citep{zha+02}. Here the focus is on the
consequences of MPs for the replenishment of the loss-cone, and the
implications for stellar populations in the Galaxy \citep{per+07},
the coalescence of binary MBHs \citep{per+07b} and for the cosmic
rates of EMRIs (Perets, Hopman \& Alexander, in prep.). 

Loss-cone replenishment by MPs can be described by the standard loss-cone
formalism (e.g. \citealt{you77}) with only few modifications \citep{per+07}.
The large size of the MPs is taken into account by decreasing the
Coulomb logarithm accordingly; the orbital averaging of phase-space
diffusion due to scattering by stars is done incoherently (sum of
squares), while for rare MPs, where there may be on average less than
one scattering events per orbital period, the averaging is done coherently
(square of sums). 

The relative contributions of relaxation by stars and relaxation by
MPs to the total loss-cone replenishment rate depend on the size of
$r_{\mathrm{crit}}$ relative to the spatial distribution of the MPs
($r_{\mathrm{crit}}$ increases with the loss-cone size, and in the
case of inspiral also with the efficiency of the dissipative process).
MPs are extended objects, which cannot survive in the strong tidal
field of the MBH (IMBHs could be the one exception). Generally, MPs
in galactic centers could also be affected by an intense central radiation
field, whether the AGN's or the stars', or by outflows associated
with accretion on a MBH. These processes introduce an inner cutoff
$r_{\mathrm{MP}}$ to the MP distribution. A plausible estimate is
$r_{\mathrm{MP}}\!\gtrsim\!{\cal O}(r_{h})$. This is the case in
the GC, where the clumpy circumnuclear gas ring lies outside the central
$1.5$ pc, on a scale comparable to $r_{h}$. The event rates of processes
such as tidal disruption of single stars ($r_{\mathrm{crit}}\!\sim\! r_{h}$)
or GW EMRI ($r_{\mathrm{crit}}\!\ll\! r_{h}$), where stellar relaxation
by itself efficiently fills the loss-cone at $r_{\mathrm{crit}}\!<\! r\!<\! r_{\mathrm{MP}}$,
will not be much enhanced by additional relaxation due to MPs (the
stellar distribution function (DF) cannot be more random than isotropic).
In contrast, the event rates of processes whose loss-cone is large,
and which would have remained empty beyond $r_{\mathrm{MP}}$ in the
absence of MPs, can be increased by orders of magnitude by the presence
of MPs. Most of the enhancement is due to MPs near $r_{\mathrm{MP}}$
\citep{per+07}.

Here we consider two processes with large loss-cones, where MPs play
an important role: the tidal disruption of stellar binaries of total
mass $M_{12}$ and semi-major axis $a_{12}$ that interact with the
MBH at a distance $r_{p}\!<\! r_{t}\!\sim\! a_{12}(\Mbh/M_{12})^{1/3}$,
leading to the capture of one star around the MBH and the ejection
of the other as a HVS \citep{hil88}, and the orbital decay of a binary
MBH of total mass $M_{12}$, mass ratio $M_{2}/M_{1}\!=\! Q\!<\!1$
and semi-major axis $a_{12}$ by interactions with stars at a distance
$r_{p}\!\lesssim\!{\cal O}(a_{12})$ (the {}``slingshot effect'')
\citep{beg+80}.

\subsection{Massive perturbers in the Galactic Center}

\label{ss:MP_GC}

\begin{figure}
\noindent \begin{centering}
\begin{tabular}{cc}
\includegraphics[width=0.5\columnwidth,keepaspectratio]{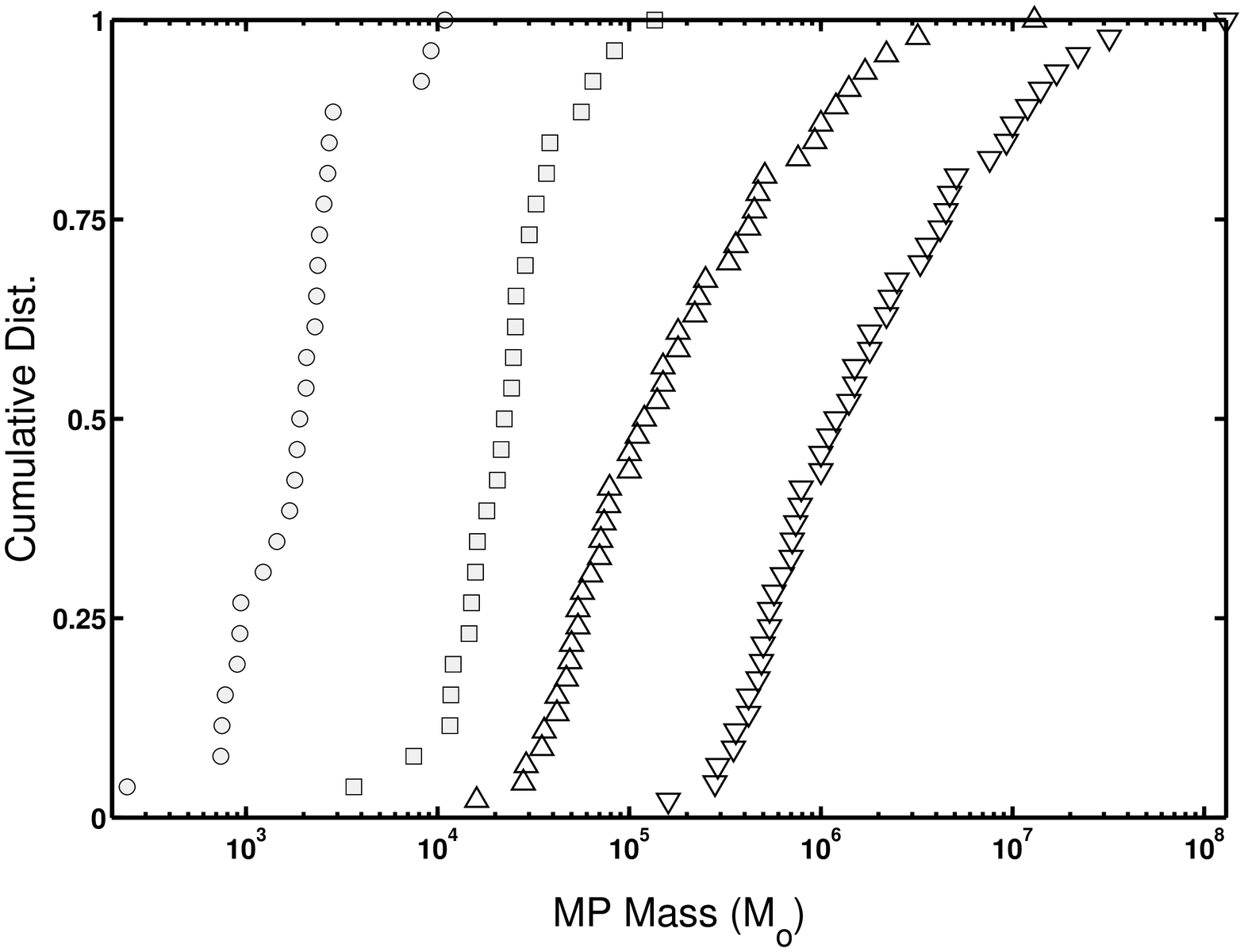} & \raisebox{0.2in}{\includegraphics[width=0.45\columnwidth,keepaspectratio]{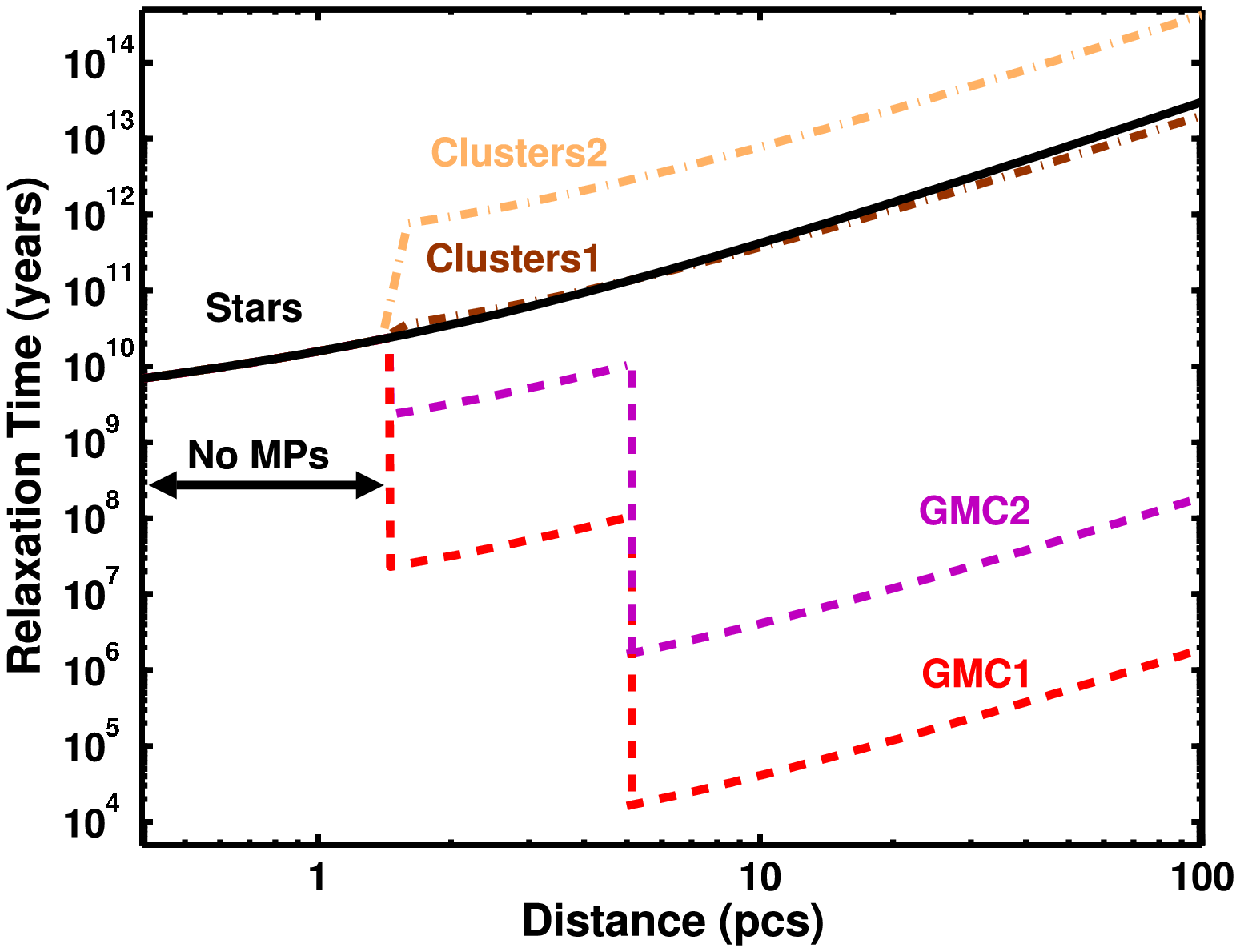}}\tabularnewline
\end{tabular}
\par\end{centering}

\caption{\label{f:GMC_MF}The observed MPs in the GC and their effect on the
relaxation time. Left: The observed mass function of molecular cloud
massive perturbers in the GC (adapted from \citet{per+07} with permission
from the \emph{Astrophysical Journal}). Lower ($\circ$) and upper
(virial) ($\square$) mass estimates for the molecular clumps in the
inner $\sim\!5$ pc, based on the molecular line observations of \citet{chr+05},
and lower ($\bigtriangleup$) and upper (virial) ($\bigtriangledown$)
mass estimates for the GMCs in the inner $\sim\!100$ pc of the Galaxy,
based on the molecular line observations of \citet{oka+01}. Right:
The relaxation time as function of distance from the Galactic MBH
due to stars alone, the upper (GMC1) and lower (GMC2) mass estimates
of the molecular clumps and GMCs and due to upper (Clusters1) and
lower (Clusters2) estimates on the number and masses of stellar clusters.
The sharp transitions at $r\!=\!1.5$ and $5$ pc are artifacts of
the non-continuous MP distribution assumed here. GMCs dominate the
relaxation in the GC.}

\end{figure}

MPs in the GC include GMCs, stellar clusters and possibly IMBHs, if
these exist. Direct observational evidence (Fig. \ref{f:GMC_MF})
indicates that the dominant MPs on the $r\!\sim\!5$--$100$ pc scale
are ${\cal O}(100)$ GMCs in the mass range $10^{4}$--$10^{8}\,\Mo$,
with rms mass of $\sim\!10^{7}\,\Mo$ and a typical size of $R_{p}\!\sim\!5$
pc (the quoted range includes an order of magnitude uncertainty in
the mass determination), \citep{oka+01,gus+04}, and on the $r\!\sim\!1.5$-$5$
pc scale, ${\cal O}(10)$ molecular clumps%
\footnote{\label{fn:clumps}The division of a quasi-continuous medium into individual
clouds is somewhat arbitrary, since several sub-clumps can be identified
as a single cloud, depending on the spatial resolution of the observations
and the adopted definition of a cloud. For a fixed total MP mass,
$M\!=\!\Mp\Np$ within a region of size $r$, the relaxation time
scales with $N_{p}$ as $T_{R}\!\propto\!(\Mp^{2}\Np)^{-1}\!=\! M^{-2}N_{p}$;
the more massive and less numerous the clouds, the shorter $T_{R}$.
The value of $T_{R}$ thus depends on the way clouds are counted.
Obviously, the statistical treatment of relaxation is valid only for
$N_{p}\!\gg\!1$ and $R_{p}\!\ll\! r$.%
} with masses in the range $10^{3}$--$10^{5}\,\Mo$, with rms mass
of $\sim\!10^{4}\,\Mo$ and a typical size of $R_{p}\!\sim\!0.25$
pc \citep{chr+05}. The $\sim\!10$ observed stellar clusters \citep{fig+99,bor+05}
may compete with stellar relaxation \citep{per+07b}. Compared to
the $\sim2\!\times\!10^{8}$ $\sim\!1\,\Mo$ stars in the central
$100$ pc \citep{fig+04}, the GMCs are expected to decrease the relaxation
time by a factor $\mu_{2}\!\sim\!50$--$5\!\times\!10^{7}$ (Eq. \ref{e:mu2MP}).
Figure (\ref{f:GMC_MF}) shows a more detailed estimate of the local
relaxation time for the various molecular cloud models, taking into
account, among other considerations, the Coulomb factors. The relaxation
time is indeed substantially decreased, by factors of $10-10^{7}$
relative to that by stars alone, depending on distance from the center,
and on the GMC mass estimates. If IMBHs do exist, then the effects
of accelerated relaxation will be even stronger than predicted here,
and probably extend all the way to the center. 

\subsection{Galactic and cosmic implications}

\label{ss:MP_imp}

With stellar relaxation alone, the empty loss-cone region of MBH-binary
interactions is large ($r_{t}\!\propto\! a_{12}$) and extends out
to $>\!100$ pc. However, the MPs that exist in the Galaxy on that
scale accelerate relaxation, efficiently fill the loss-cone, and thus
increase the binary disruption rate by several orders of magnitude,
making binary disruptions dynamically and observationally relevant
\citep{per+07}. Such events, which result in the energetic ejection
of one star, and the capture of the other on a close orbit around
the MBH, have various possible implications. Disruptions of binaries
by the Galactic MBH \citep{hil88,yuq+03,gua+05,bro+06c} were suggested
to be the origin of the hyper-velocity B-stars%
\footnote{HVS candidates are chosen for spectroscopy by color, to maximize the
contrast against the halo population, and so are pre-selected to have
B-type spectra \citep[e.g.][]{bro+06b}.%
} ($v\gtrsim\!500\,\mathrm{km\, s^{-1}}$), observed tens of kpc away
from the GC \citep{hir+05,bro+05,ede+06,bro+06a}, and the origin
of the puzzling {}``S-stars'' \citep{gou+03,gin+06}, a cluster
of $\sim\!10$--$30$ main-sequence B-stars ($4\,\Mo\!\lesssim\!\Ms\!\lesssim\!15\,\Mo$,
main sequence lifespan $t_{\star}\!\sim\mathrm{few}\!\times\!10^{7}$--$\mathrm{few\!\times\!10^{8}}$
yr) on random tight orbits around the MBH in the central $\mathrm{few\times0.01}$
pc \citep{eis+05,ghe+05}. Compact objects captured this way could
eventually become zero-eccentricity GW sources \citep{mil+05}, in
contrast to high-eccentricity sources typical of single-star inspiral
\citep{hop+05}. These two classes of sources are expected to emit
very different gravitational wave-forms.

\begin{figure}
\noindent \begin{centering}
\begin{tabular}{cc}
\includegraphics[width=0.45\columnwidth]{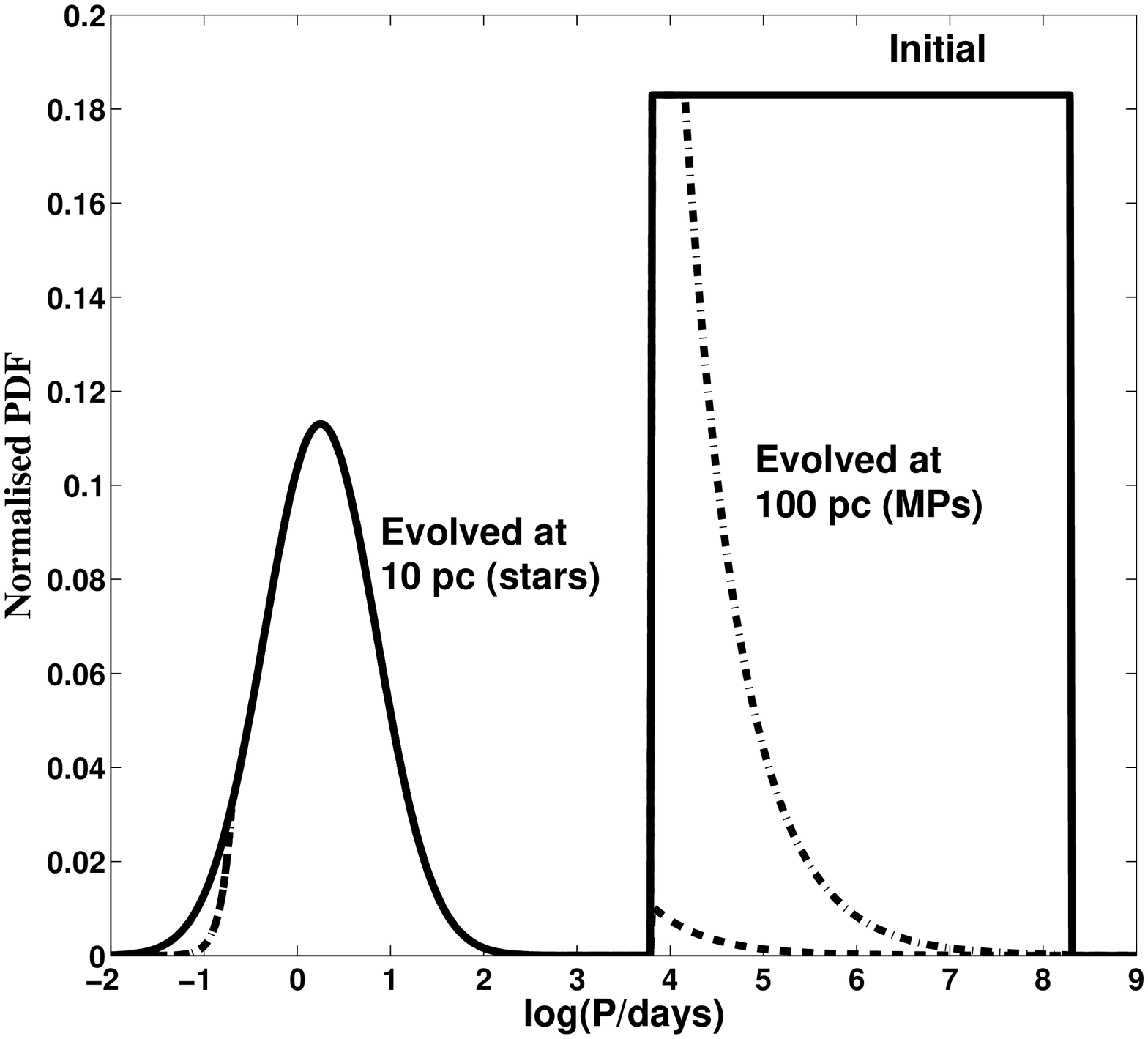} & \includegraphics[width=0.45\columnwidth]{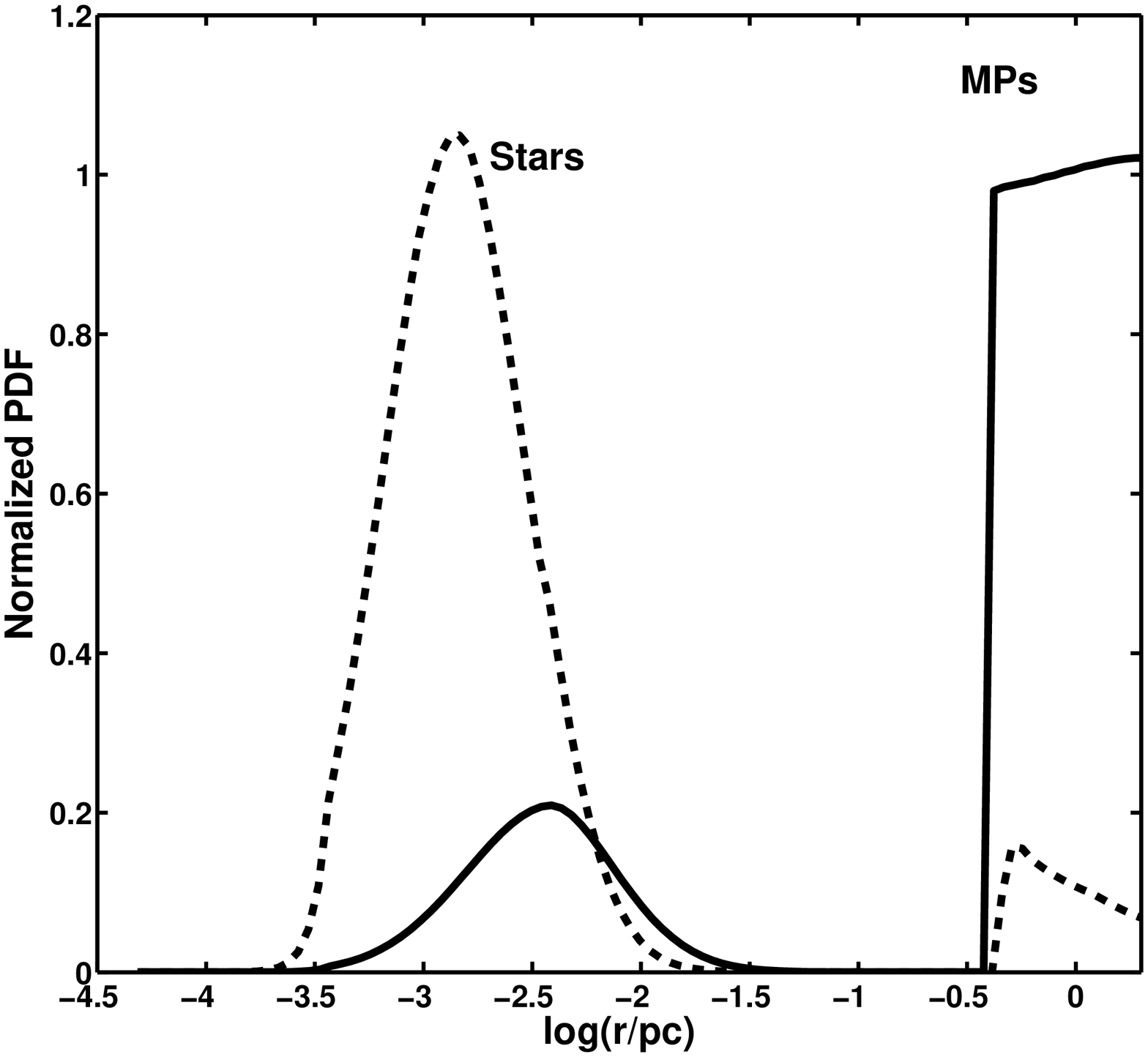}\tabularnewline
\end{tabular}
\par\end{centering}

\caption{\label{f:map}A schematic representation of the mapping of the initial
binary period distribution to the semi-major axis of the tidally captured
star. Left: The bimodal initial period distribution for old white
dwarf / main sequence binaries (adapted from \citealt{wil+04b}),
and its subsequent evolution due to GW coalescence (for the shortest
periods) and to slow evaporation by field stars at 100 pc and faster
evaporation at 10 pc. Right: the resulting semi-major axis distribution
of the captured stars, due to scattering by stars, which occurs on
the ${\cal O}(10\,\mathrm{pc})$ scale, and due to scattering by MPs,
which occurs on the ${\cal O}(100\,\mathrm{pc})$ scale.}

\end{figure}

Dynamical arguments and simulations show that on average, $\sim0.75$
of MBH-binary encounters lead to capture, and that the mean semi-major
axis of the captured star is related to that of the original binary
by \citep{hil88,hil91} 

\begin{equation}
\left\langle a\right\rangle \sim\left(\Mbh/M_{12}\right)^{2/3}a_{12}\,,\label{e:acapture}\end{equation}
which implies a very high initial eccentricity, $1-e\!=\! r_{t}/\left\langle a\right\rangle \!=\!(M_{12}/\Mbh)^{1/3}\!\sim\!{\cal O}(0.01)$.
The tidal capture process can be viewed as a mapping between the properties
of field binaries far from the MBH, and the orbital properties of
the captured stars: wide binaries result in wide captured orbits,
and vice versa (Fig. \ref{f:map}). The mean velocity of the ejected
star at infinity (neglecting the galactic potential) is \begin{equation}
\left\langle v_{\infty}^{2}\right\rangle \sim\sqrt{2}GM_{12}^{2/3}\Mbh^{1/3}/a_{12}\,.\end{equation}
This translates, for example, to $v_{\infty}\!\sim\!2000\,\mathrm{km\, s^{-1}}$
for a $2\!\times\!4\,\Mo$ B-star binary with $a_{12}\!=\!0.2$ AU,
well above the escape velocity from the Galaxy. 

\begin{figure}
\noindent \begin{centering}
\includegraphics[width=0.85\columnwidth]{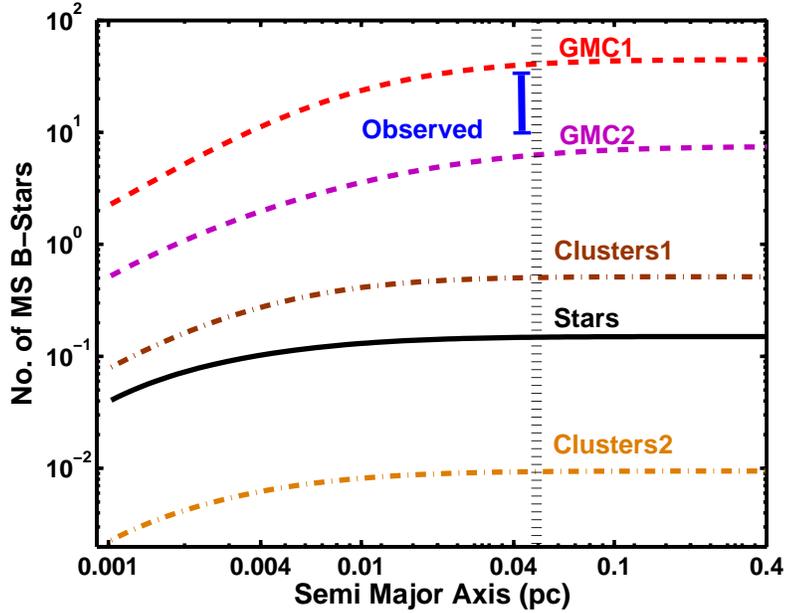}-
\par\end{centering}

\caption{\label{f:Sstars} A comparison between the cumulative number of S-stars
(main sequence B stars) observed orbiting the Galactic MBH on randomly
oriented orbits (vertical bar), and the predicted number captured
by 3-body tidal interactions of the MBH with binaries deflected to
the center by massive perturbers, for different massive perturbers
models \citep{per+07}. The observed extent of the S-star cluster
is indicated by the vertical hashed line. }

\end{figure}

Of particular interest is the connection between the HVSs and the
S-stars that is implied by the binary tidal disruption scenario. The
stellar binary mass ratio distribution is peaked around $\sim\!1$
\citep{duq+91,kob+06}, and so the observed similarity in the spectral
type of the S-stars and HVSs is consistent with this scenario. Figure
(\ref{f:Sstars}) shows an estimate of the number of tidally captured
S-stars for different MP populations \citep{per+07}, based on the
observed orbital properties of young massive binaries ($a\!\sim\!0.20_{-0.15}^{+0.60}$
AU) and their fraction among young massive stars in the field ($f_{2}\!\sim\!0.75$)
\citep{kob+06}, and on a model for the stellar density distribution
in the inner $\sim\!100$ pc of the Galaxy (isothermal, normalized
by the observations of \citealt{gen+03a}) and a mass function model
(continuous star formation with a universal IMF, \citealt{fig+04},
see also Fig. \ref{f:HBRC}). The typical binary was modeled as a
$2\times7.5\,\Mo$ binary (main sequence B-stars with a lifespan of
$t_{\star}\!\sim\!5\times10^{7}$ yr). Dynamical evaporation is negligible
for such short-lived binaries. The steady state number of captured
S-stars is then $\left\langle \Ns\right\rangle \!=\!\Gamma_{lc}t_{\star}$.
Figure (\ref{f:Sstars}) shows that with stellar relaxation alone,
tidal capture cannot explain the S-star population. However, relaxation
by GMCs is consistent with the observed number of S-stars, as well
as with the spatial extent of the cluster of $\sim0.04$ pc, which
reflects the hardness of young massive binaries in the field (Eq.
\ref{e:acapture}). It is also consistent with the fact that the S-cluster
does not include any star earlier then O8V/B0V. Such short-lived binaries
are very rare in the field, and their mean number in the S-cluster
is predicted to be $\left\langle \Ns\right\rangle \!<\!1$. 

The MP-induced binary tidal disruption scenario also predicts that
there should be $10$--$50$ hyper-velocity $\sim\!4\,\Mo$ B-stars
at distances between $20$ and $120$ kpc from the GC. This is consistent
with the total number of $43\!\pm\!31$ extrapolated by \citet{bro+06a},
based on the HVSs detected at these distances in their field of search.
The tidal disruption scenario predicts an isotropic distribution of
HVSs around the GC, and a random ejection history, in contrast to
models where the ejection is related to a discrete binary MBH merger
event \citep{yuq+03,haa+06,bau+06,lev06c}. The HVSs observed to date
are consistent with an isotropic HVS distribution uniformly distributed
in ejection time \citep{bro+06a} and thus support the tidal disruption
scenario.

The tidal disruption scenario can naturally explain many of the properties
of the S-stars and HVSs, but it has two potential flaws. (1) The predicted
high eccentricities of the captured stars are larger than those observed
for a few of the S-stars ($e\!\sim\!0.4$, \citealt{eis+05}). However,
the low observed eccentricities are expected to evolve after the capture
by rapid resonant relaxation (\S \ref{s:RR}). (2) The lifespan of
the most massive and shortest lived S-stars ($t_{\star}\!\sim\!\mathrm{2\!\times\!10^{7}}$
yr) is shorter by a factor $\lesssim\!10$ than the MP-accelerated
relaxation time in the inner $\sim\!5$ pc (Fig. \ref{f:GMC_MF}),
where a substantial fraction of the binaries are scattered from. Thus
if a binary in those regions starts on a near-circular orbit, MP-induced
relaxation is not fast enough to scatter it to a $J\!<\! J_{lc}$
orbit (Eq. \ref{e:TJ}) within its lifetime. However, as the timescale
discrepancy is not large, and as it affects only the most massive
binaries in the central few pc, where the determination of $T_{R}$
is ambiguous (see footnote \vpageref{fn:clumps}), this does not appear
to be a fatal flaw of this scenario. It does however highlight the
importance of observationally quantifying the relaxation time in the
GC and the distribution and properties of the field binaries. 

Low-mass binaries are also deflected to the MBH by MPs and tidally
disrupted at rates as high as $\sim\!10^{-4}\,\mathrm{yr^{-1}}$ \citep{per+07}.
Neither the faint captured low-mass stars nor the late-type HVSs are
detectable at this time. However, the captured stars affect the inner
cusp dynamics in a way that may have implications for cosmic GW EMRI
events. Binary disruption is effectively a local {}``source term''
that modifies the flow of stars in phase space (cf Eq. \ref{e:FPeq}),
setting a diverging flow into the MBH and away from it, which modifies
the steady state spatial distribution. Detailed calculations, which
take into account the period distribution of low-mass binaries and
the effects of binary evaporation, indicate that MP-induced tidal
captures of white dwarfs close to the MBH efficiently competes against
mass-segregation, which tends to lower the density of the low-mass
white dwarfs there and raise the density of massive stellar BHs (\S
\ref{s:MSeg}, Fig. \ref{f:GCmseg}). As a result, the cosmic rate
of GW EMRI events involving white dwarfs is predicted to be at least
comparable to that involving stellar BHs (Perets et al. 2007, in prep.).

The proximity of the GC allows GW bursts from the fly-by of stars
near the MBH to be detected \citep{rub+06}. MP-induced tidal binary
disruptions increase the stellar density close to the MBH and therefore
the rate of GW bursts increases significantly. In particular, the
rate of GW bursts from white dwarfs increases from $\sim\!0.1\,\mathrm{yr^{-1}}$
\citep{hop+07} to a detectable rate of $\sim\!2\,\mathrm{yr^{-1}}$
(Perets et al. 2007, in prep.).

\begin{figure}
\noindent \begin{centering}
\begin{tabular}{cc}
\includegraphics[width=0.5\columnwidth,keepaspectratio]{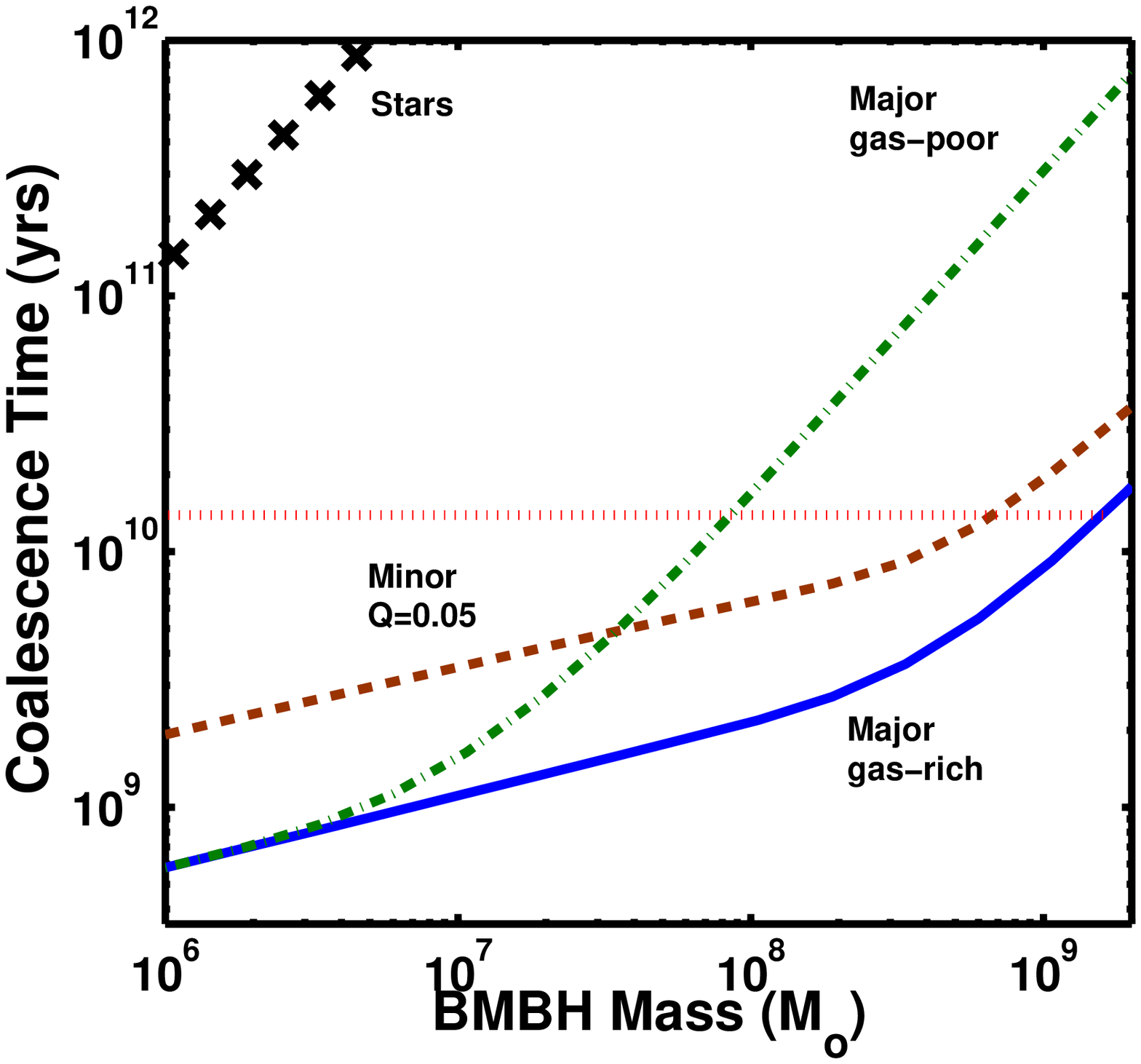} & \includegraphics[width=0.5\columnwidth,keepaspectratio]{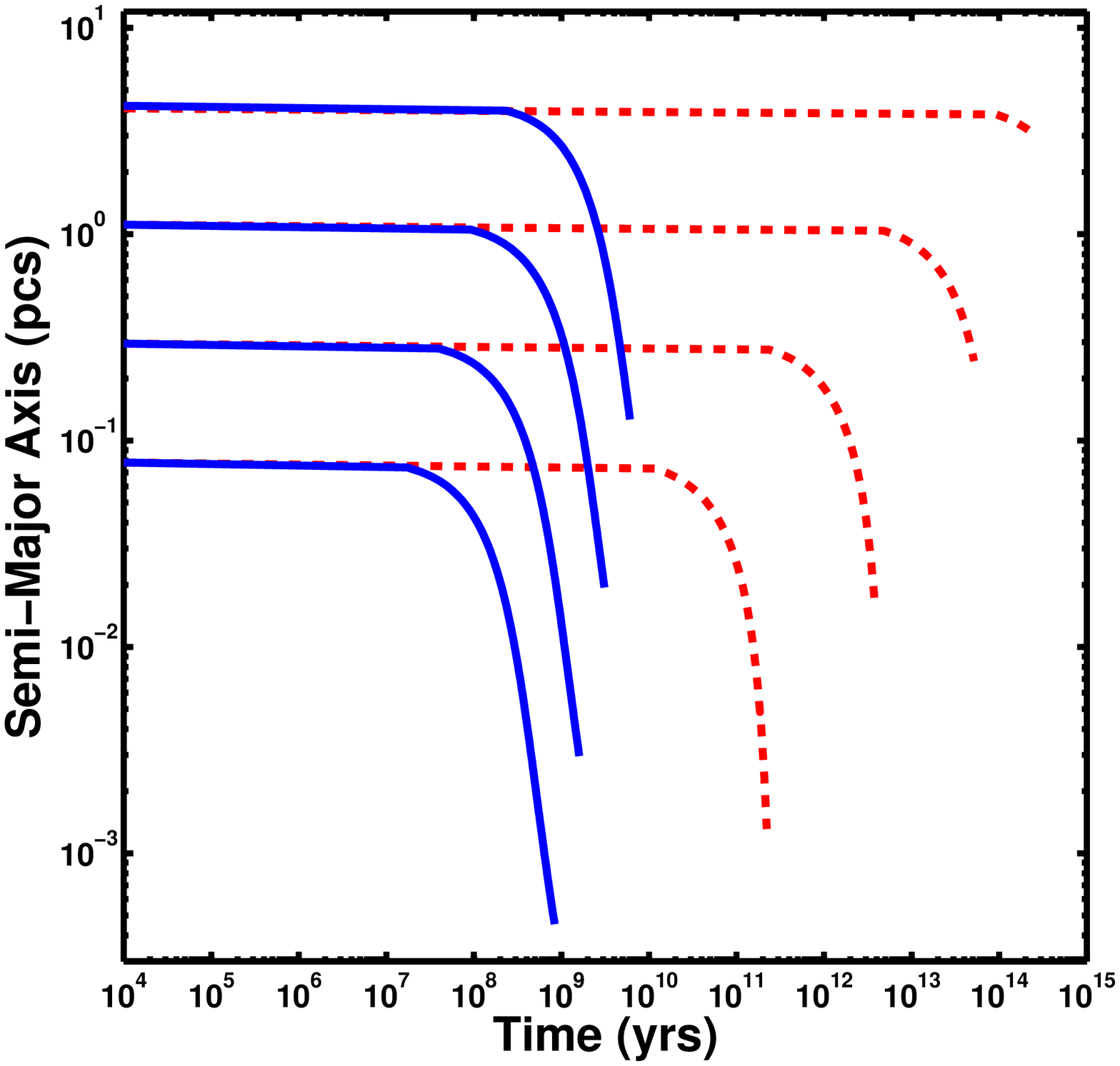}\tabularnewline
\end{tabular}
\par\end{centering}

\caption{\label{f:BMBH} Accelerated binary MBH mergers in the presence of
MPs \citep{per+07b}. Left: The time to coalescence as function of
binary MBH mass, for different merger scenarios distinguished by the
mass ratio $Q$ between the two MBHs and the MP contents of host galaxies.
The age of the universe is indicated by the dotted horizontal line.
Stellar relaxation alone cannot supply a high enough rate of stars
for the slingshot mechanism to complete the merger within a Hubble
time. However, in minor mergers ($Q\!=\!0.05$) major gas-rich mergers
($Q\!=\!1$) with MPs merger is possible within a Hubble time for
all but the most massive MBHs. Right: The evolution of the binary
MBH semi-major axis as function of time for major mergers ($Q\!=\!1$)
in the presence of MPs (solid line) and stellar relaxation alone (dashed
line), for binary MBH masses of $10^{6}$, $10^{7}$, $10^{8}$ and
$10^{9}\,\Mo$ (from bottom up).}

\end{figure}

Binary MBHs form in the aftermath of galactic mergers, when the two
MBHs sink by dynamical friction to the center of the merged galaxy.
Once the binary hardens, the orbital decay continues by 3-body interactions
with stars that are deflected to the center and extract energy from
the binary, until the orbit becomes tight enough for efficient GW
emission, which rapidly leads to coalescence. Simulations show that
when the loss-cone is replenished by stellar relaxation alone, the
interaction rate is too slow for the binary MBH to coalesce within
a Hubble time (e.g. \citealt{ber+05}; see review by \citealt{mer+05b};
Fig. \ref{f:BMBH}). This {}``last parsec stalling problem'' appears
to contradict the circumstantial evidence that most galactic nuclei
contain only a single MBH \citep{ber+06,mer+05b}, and furthermore
implies few such very strong GW sources for LISA. One route%
\footnote{Other possible routes are by interactions with gas in {}``wet mergers''
\citep{iva+99,esc+05,dot+06b}, by interactions with a third MBH \citep{mak+94,bla+02,iwa+06},
or by accelerated loss-cone replenishment in a non-axisymmetric potential,
\citep{yuq02,ber+06}, or in a steep cusp \citep{zie06a}. %
} for resolving the stalling problem is by accelerated MP-induced loss-cone
replenishment \citep{per+07b} 

Figure (\ref{f:BMBH}) shows the time to coalescence, as function
of the binary MBH mass, for different merger and MP scenarios, based
on a combination of extrapolation of the Galactic MP population to
early type galaxies, on extra-galactic observations of molecular gas
in galactic centers, and on results from galactic merger simulations.
The results show that MPs allow binary MBHs in gas-rich galaxies to
coalesce within a Hubble time over nearly the entire range of $M_{12}$.
The situation with respect to gas-poor galaxies is less clear, since
it is harder to model reliably the MPs there (probably clusters rather
than GMCs). However, even for such galaxies, MPs allow coalescence
within a Hubble time up to masses of $M_{12}\lesssim\!10^{8}\,\Mo$. 

Efficient binary MBH coalescence by MPs has various implications.
It increases the cosmic rate of GW events from MBH-MBH mergers, it
increase the {}``mass deficit'' in the galactic core (the stellar
mass ejected from the core by the slingshot effect) \citep{mil+02,rav+02,gra04,fer+06},
it leads to the ejection of hyper-velocity stars to the inter-galactic
space, but it suppress the formation of triple MBH systems and the
ejection of MBHs into intergalactic space \citep{sas+74,bla+02,hof+06b,iwa+06}. 

\section{Strong mass segregation}

\label{s:MSeg}

\subsection{The Bahcall-Wolf solution of moderate mass-segregation}

\label{ss:BW77}

The 2-body relaxation timescale around the Galactic MBH, $T_{R}\sim{\cal O}(1\,\mathrm{Gyr})$,
is short enough for the old stellar population there to relax to a
universal steady-state configuration, independently of the initial
conditions. This configuration was investigated by \citet{bah+76,bah+77}.
The Bahcall-Wolf solution predicts that in the Keplerian potential
near a MBH, stars of mass $\Ms$ in a multi-mass population ,$M_{1}\!<\!\Ms\!<\! M_{2}$,
have a DF that is approximately a power-law of the specific orbital
energy $\epsilon$, $f_{M}(\epsilon)\!\propto\!\epsilon^{p_{M}}$,
where $p_{M}\!\propto\!\Ms$ with a proportionality constant $p_{M}/\Ms\!\simeq\!1/(4M_{2})$.
In a Keplerian potential, this DF corresponds to a density cusp $n_{M}(r)\!\propto\! r^{-\alpha_{M}},$
where $\alpha_{M}\!=\!3/2+p_{M}$. Elementary considerations show
that $\alpha\!=\!7/4$ ($p\!=\!1/4$) for a single mass population
\citep[e.g. ][$\S$ 8.4-7]{bin+87}. This follows from the conservation
of the orbital energy that is extracted from stars that are scattered
into the MBH, and transferred outward by the ambient scattering stars
in a steady-state, distance-independent current, $\mathrm{d}E(r)/\mathrm{d}t\!\sim\! N_{\star}(<\! r)E_{\star}(r)/T_{R}(r)\!\sim\! r^{7/2-2\alpha}\!=\!\mathrm{const}$
(using the relations $N_{\star}(<r)\!\propto\! r^{3-\alpha}$, $E_{\star}\!\sim\! r^{-1}$
and $T_{R}\!\propto\! r^{\alpha-3/2}$, \S \ref{ss:GCdyn}). The
Bahcall-Wolf solution reproduces this result for a single mass population,
and predicts that it should apply also to the heaviest stars in a
multi-mass population. The Bahcall-Wolf solution thus implies that
at most $\Delta\alpha\!=\!1/4$ between the lightest and heaviest
stars in the population. The predicted degree of segregation is moderate.

\begin{figure}
\noindent \begin{centering}
\begin{tabular}{cc}
\includegraphics[width=0.5\columnwidth,keepaspectratio]{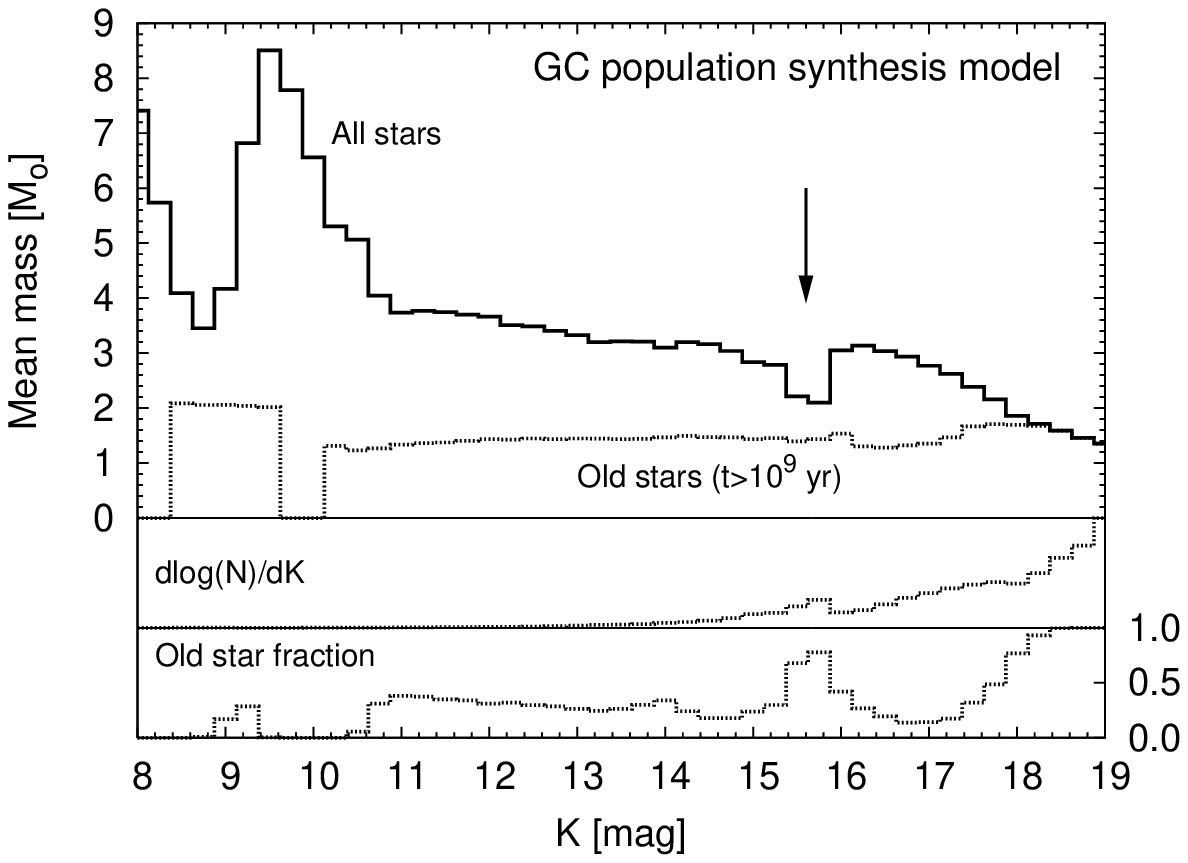} & \includegraphics[width=0.5\columnwidth,keepaspectratio]{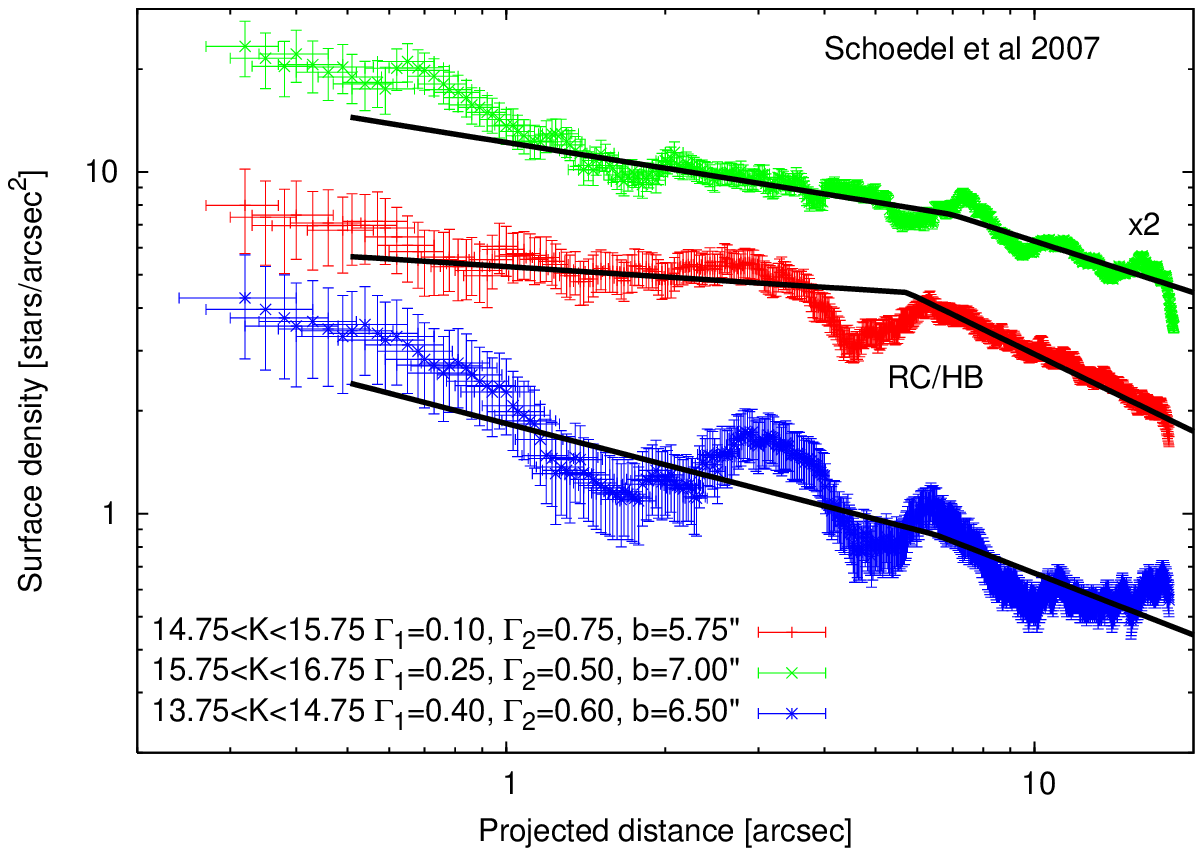}\tabularnewline
\end{tabular}
\par\end{centering}

\caption{\label{f:HBRC} Left: A theoretical population model for the central
few pc of the GC (\citealt{ale+99a,ale05}), assuming continuous star
formation over the past 10 Gyr \citep{fig+04} with a {}``universal''
IMF \citep{mil+79}. Bottom panel: The fraction of old stars (defined
here as stars with main-sequence lifespan of $>1$ Gyr) in the population,
as function of the $K$-band magnitude in the GC (for $DM\!+\! A_{K}\!=\!17.2$
mag, \citealt{eis+05}). Middle panel: The $K$-band luminosity function.
Top panel: The mean mass of all stars and of the old stars only, as
function of the $K$-band magnitude. The concentration of old Red
Clump / horizontal branch giants around $K\!\sim\!15.5$ is clearly
seen as an excess in the luminosity function, as an increase the fraction
of old stars and as a decrease in the mean stellar mass relative to
stars both immediately brighter and fainter \citep{sch+07}. Right:
The observed azimuthally-averaged stellar surface number density around
the Galactic MBH as function of projected angular distance, $\Sigma(R)$
($R\!=\!1"$ corresponds to $\sim\!0.04$ pc in the GC) in 3 adjacent
$K$-magnitude bins, centered around the bin associated with the Red
Clump giants ($14.75\!<\! K\!<\!15.75$), with broken power-law fits
$\Sigma\!\propto\! R^{-\Gamma}$ (\citealt{sch+07}, adapted with
permission from \emph{Astronomy and Astrophysics}). Top: $15.75\!<\! K\!<\!16.75$
(density multiplied by $2$ for display purposes). Middle: $14.75\!<\! K\!<\!15.75$
(the Red Clump / horizontal branch range). Bottom: $13.75\!<\! K\!<\!14.75$.}

\end{figure}

Theoretical considerations, results from dynamical simulations and
GC observations, hint that the moderate segregation solution should
not and does not always hold, even in relaxed systems. As formulated,
the solution depends only on the stellar masses, but not on the mass
function. However, this cannot apply generally, since in the limit
where the massive objects are very rare, they are expected to sink
efficiently to the center by dynamical friction, and create a cusp
much steeper than $\alpha\!=\!7/4$. As shown below (\S \ref{ss:mu2}),
models of the present-day mass function in the central few pc of the
GC suggest that the massive objects are relatively rare. Dynamical
simulations of mass segregation in the GC based on such a mass function
(Fig. \ref{f:GCmseg}) indeed show steep cusps ($\alpha\!>\!2$) for
the heaviest masses. Finally, the observed surface density distribution
of GC stars in the magnitude bin $14.75\!<\! K\!<\!15.75$, which
corresponds to the low-mass ($0.5\!\lesssim\!\Ms\!\lesssim\!2\,\Mo$)
Red Clump / horizontal branch giants (Fig. \ref{f:HBRC}), is substantially
flatter than that of the higher-mass giants ($\Ms\!\sim\!3\,\Mo$)
that populate the adjacent bins of brighter and fainter magnitudes
(Fig. \ref{f:HBRC}; \citealt{sch+07}). The sign of this trend is
as expected for mass segregation, but the size of the effect is much
larger than predicted by the Bahcall-Wolf moderate segregation solution.
However, it can be explained in terms of mass-segregation if $\Delta\alpha\!\gtrsim\!1$
\citep{lev06b}. While none of these hints for strong mass-segregation
is decisive in itself, and other explanations are possible, taken
together they motivate a re-examination of the mass-segregation solution
in a relaxed system. 

\subsection{The relaxational self-coupling parameter}

\label{ss:mu2}

Assume for simplicity a stellar system with a two-mass population
of light stars of mass $M_{L}$, total initial number $N_{L}$ and
local density $n_{L}(r)$ and heavy stars of mass $M_{H}$, total
initial number $N_{H}$ and local density $n_{H}(r)$. The self interaction
rate is then $\Gamma_{LL}\!\propto\! n_{L}M_{L}^{2}/v^{3}$ for the
light stars and $\Gamma_{HH}\!\propto\! n_{H}M_{H}^{2}/v^{3}$ for
the heavy stars (\S \ref{s:MP}). In the limit where the heavy stars
are test particles ($n_{H}/n_{L}\!\ll\! M_{L}^{2}/M_{H}^{2}$, or
equivalently $\Gamma_{HH}/\Gamma_{LL}\!\ll\!1$),  the heavy stars
interact mostly with the light ones, lose energy and sink to the center
by dynamical friction. Conversely, in the limit $\Gamma_{HH}/\Gamma_{LL}\!\gg\!1$,
the heavy stars interact mostly with each other, effectively decouple
from the light stars and establish an $\alpha\!=\!7/4$ cusp typical
of a single mass population. This suggests that the \emph{global}
relaxational self-coupling parameter (cf Eq. \ref{e:mu2MP}), defined
as

\begin{equation}
\mu_{2}\!\equiv\! N_{H}M_{H}^{2}/N_{L}M_{L}^{2}\,,\end{equation}
can be used to determine whether the system settles into the moderate
(Bahcall-Wolf) mass-segregation solution ($\mu_{2}\!>\!1$) or the
strong mass-segregation solution ($\mu_{2}\!<\!1$). This hypothesis
is borne out by the numerical results presented below%
\footnote{In the limit $M_{H}/M_{L}\!\gg\!1$, it may be necessary to take explicitly
into account the the dynamical friction timescale in order to obtain
a more accurate segregation criterion. Here $\mu_{2}$ is adopted
for its simplicity.%
} (Alexander \& Hopman 2007, in prep.; \S \ref{ss:FP}). For a continuous
mass distribution, $\mu_{2}$ can be generalized to \begin{equation}
\mu_{2}\!\equiv\!\left.\int_{M_{0}}^{M_{2}}\Ms^{2}(\mathrm{d}N/\mathrm{d}\Ms)\mathrm{d}\Ms\right/\int_{M_{1}}^{M_{0}}\Ms^{2}(\mathrm{d}N/\mathrm{d}\Ms)\mathrm{d}\Ms\,,\end{equation}
the ratio between the 2nd moments of the mass distribution of the
heavy ($\Ms\!>\! M_{0}$) and light ($\Ms\!<\! M_{0}$) stars, for
some suitable choice of the light/heavy boundary mass $M_{0}$.

\begin{figure}
\noindent \begin{centering}
\includegraphics[width=0.9\columnwidth]{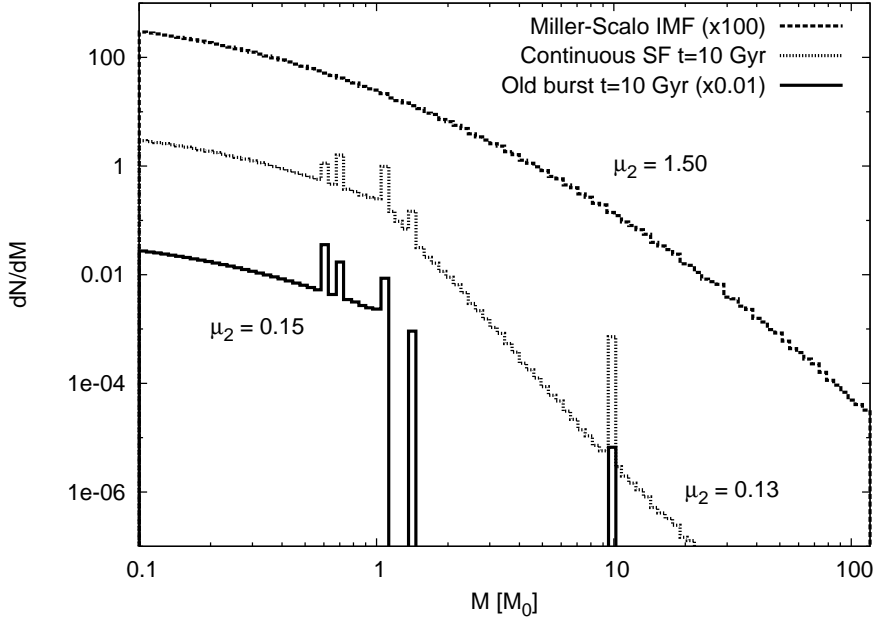}
\par\end{centering}

\caption{\label{f:mu2} The predicted values of the global relaxational self-coupling
parameter $\mu_{2}$ for a {}``universal'' \citet{mil+79} IMF (top
line, shifted by $\times100$ for display purposes), an evolved mass
function assuming continuous stars formation over 10 Gyr (middle line),
and an evolved star formation burst 10 Gyr old (bottom line, shifted
by $\times0.01$ for display purposes) (Alexander \& Hopman 2007,
in prep.). The mass functions of the old populations develop excesses
in the $\sim\!0.6$--$1.4\,\Mo$ range due to the accumulation of
white dwarfs and neutron stars, and in the $\sim\!10\,\Mo$ range
due to the accumulation of stellar black holes (here represented by
a simplified discrete mass spectrum, see \citealt{ale05}, table 2.1). }

\end{figure}

The value of $\mu_{2}$ depends on the population's present-day mass
function. So-called universal initial mass functions (IMFs), which
extend all the way from the brown dwarf boundary $M_{1}\!\sim\!0.1\,\Mo$
to $M_{2}\!\gtrsim\!100\,\Mo$ (e.g. the \citet{sal55} IMF, and its
subsequent refinements, the \citet{mil+79} and \citet{kro01} IMFs),
result in evolved populations, old star-bursts or continuously star
forming populations, that naturally separate into two mass scales,
the ${\cal O}(1\,\Mo)$ scale of low-mass main-sequence dwarfs, white
dwarfs and neutrons stars, and the ${\cal O}(10\,\Mo)$ scale of stellar
black holes, and typically have $\mu_{2}\!<\!1$ (Fig. \ref{f:mu2}).
Such evolved populations are thus well-approximated by the simple
2-mass population model. In particular, the volume-averaged stellar
population in the central few pc of the GC is reasonably well approximated
by a 10 Gyr old, continuously star-forming population with a universal
IMF%
\footnote{Note that recent analysis of late type giants in the GC suggests that
the IMF in the inner $\sim\!1$ pc of the GC could typically be a
flat $\gamma\!\sim\!0.85$ power-law \citep{man+07}. This would imply
$\mu_{2}\!\gg\!1$ in the inner $\sim\!1$ pc, possibly a volume-averaged
$\mu_{2}\!>\!1$ in the inner few pc (the {}``collection basin''
for stellar BHs, \citealt{mir+00}), and hence moderate segregation. %
} (\citealt{ale+99a}; Fig. \ref{f:HBRC}). Generally, 10 Gyr old,
continuously star-forming populations with a power-law IMF, $\mathrm{d}N/\mathrm{d}\Ms\!\propto\!\Ms^{-\gamma}$,
have $\mu_{2}\!<\!1$ for $\gamma\!\gtrsim\!2$, and $\mu_{2}\!>\!1$
for $\gamma\!\lesssim\!2$. Since the critical value $\gamma\!=\!2$
is close to the generic Salpeter index $\gamma\!=\!2.35$, it is quite
possible that both the moderate and strong segregation solutions are
realized around galactic MBHs, depending on the system-to-system scatter
in the IMF (and perhaps also realized in clusters around IMBHs, if
such exist).

\subsection{Solutions of the Fokker-Planck energy equation}

\label{ss:FP}

The steady state configuration of stars around a MBH can be described
in terms of the diffusion of stars in phase space, from an infinite
reservoir of unbound stars with a given mass function (the host galaxy,
far from the MBH), to an absorbing boundary at high energy where stars
are destroyed (the MBH event horizon, tidal disruption radius, or
collisional destruction radius).

\citet{bah+76,bah+77} simplified the full Fokker-Planck treatment
in ($E,J$) phase space by integrating over $J$ so as to reduce it
to $E$ only, by assuming a Keplerian potential, and by recasting
it in the form of a particle conservation equation. In dimensionless
form these can be written as \citep{hop+06b}

\begin{figure}
\noindent \begin{raggedright}
\begin{tabular}{cc}
\raisebox{1.9in}{\includegraphics[width=0.375\columnwidth,keepaspectratio,angle=270]{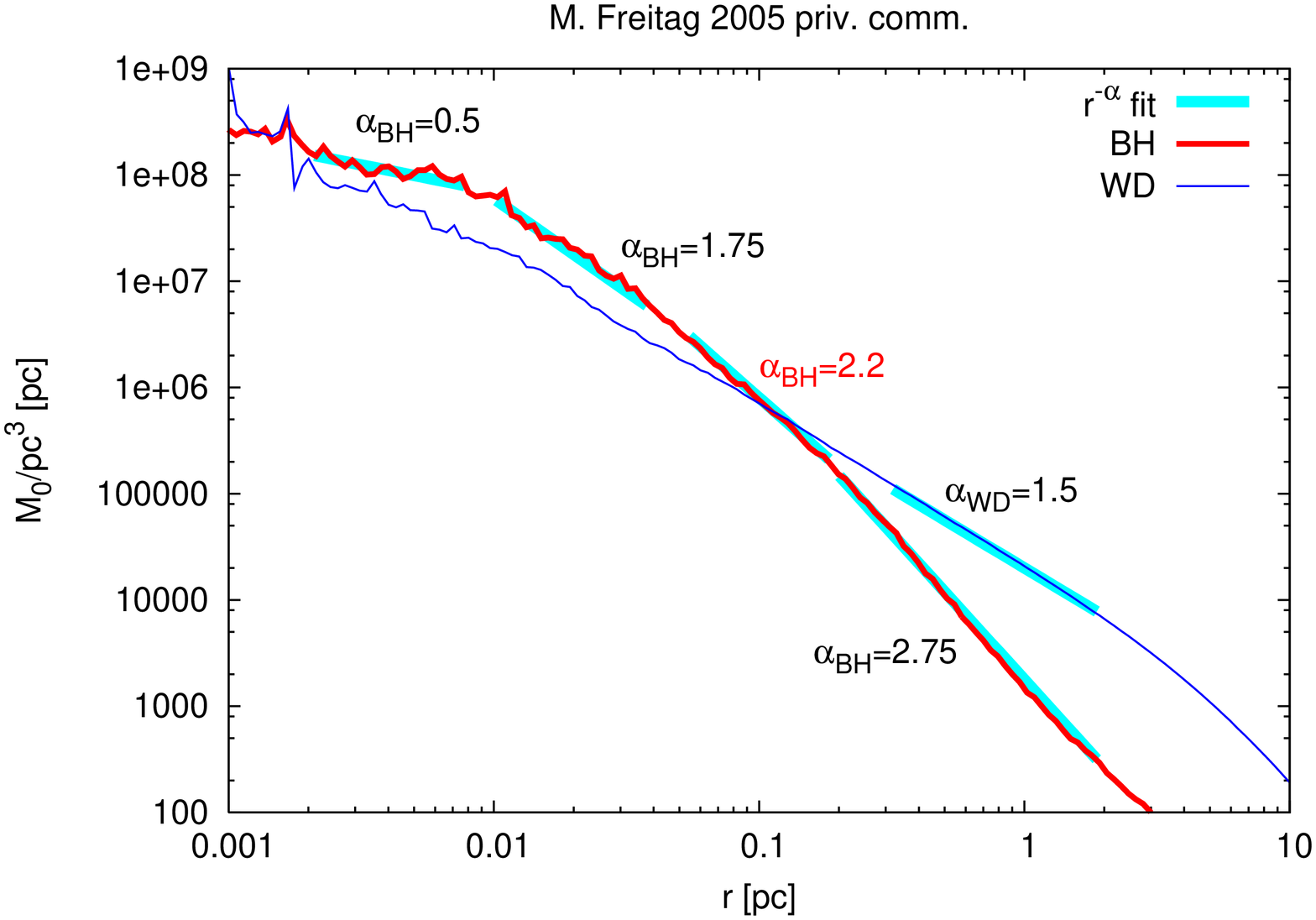}} & \includegraphics[width=0.475\columnwidth,keepaspectratio]{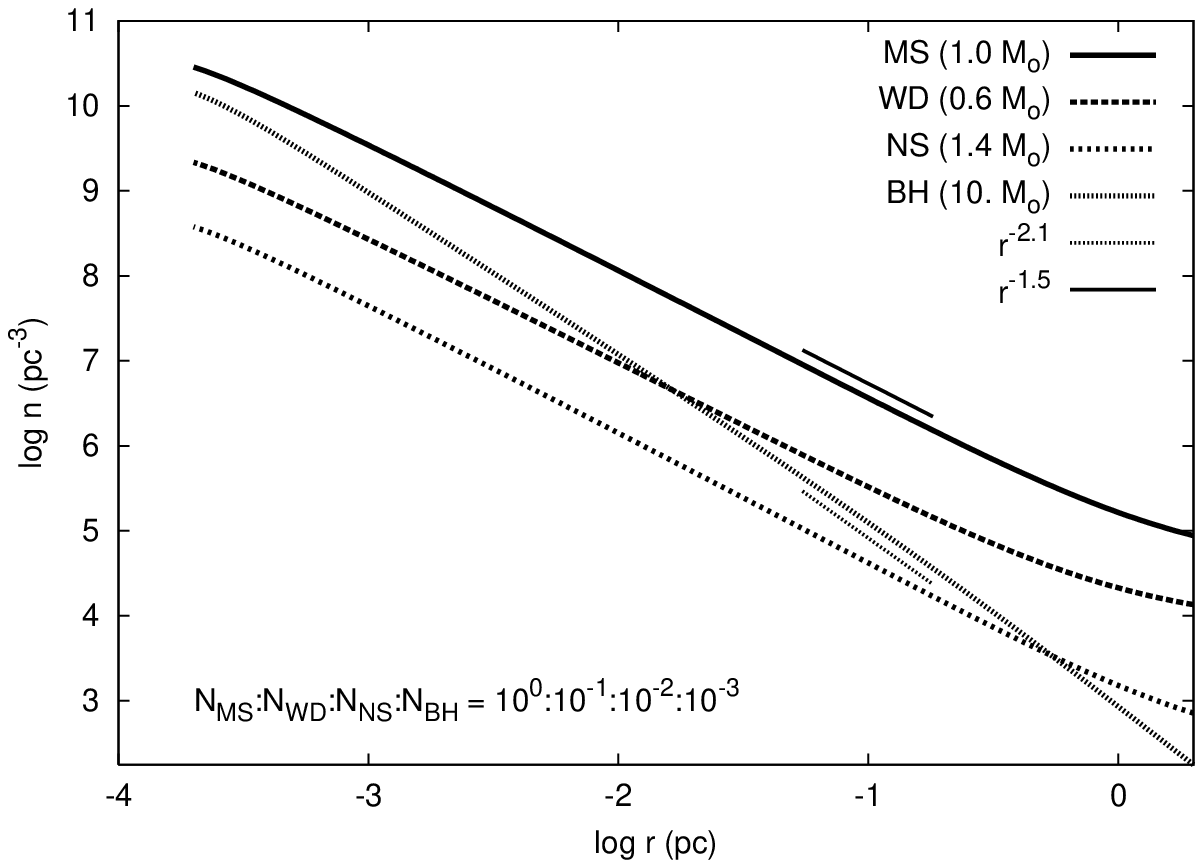}\tabularnewline
\end{tabular}
\par\end{raggedright}

\caption{\label{f:GCmseg} Numerical models of the mass distribution in the
GC showing strong segregation. Right: The density distribution of
$10\,\Mo$ stellar BHs and $0.7\,\Mo$ white dwarfs at 10 Gyr in an
approximate model of the GC, with the evolved universal IMF of Fig.
(\ref{f:BMBH}), derived by M. Freitag (priv. comm., reproduced here
with permission. See also \citealt{fre+06}), using an implementation
of the H\'enon method \citep{fre+02}. The low-mass white dwarfs
settle into a $\alpha_{L}\!\simeq\!1.5$ power-law cusp. The distribution
of massive stellar BHs can be approximated by a piece-wise broken
power-law, with $\alpha_{H}\!\sim\!2.2$ at $r\!\sim\!0.1$ pc. Left:
The space density of a simplified 4-component population model for
the GC, as given by the solution of the Fokker-Planck equations with
a loss-cone term (adapted from \citealt{hop+06b}, with permission
from the \emph{Astrophysical Journal}). The logarithmic slope of the
density cusp of the stellar BHs at $r\!=\!0.1$ pc is $\alpha\!\simeq\!2.1$,
as compared to $\alpha\!\simeq\!1.5$ for the lighter species.}

\end{figure}

\begin{figure}
\noindent \begin{centering}
\begin{tabular}{c}
\includegraphics[width=0.95\columnwidth,keepaspectratio]{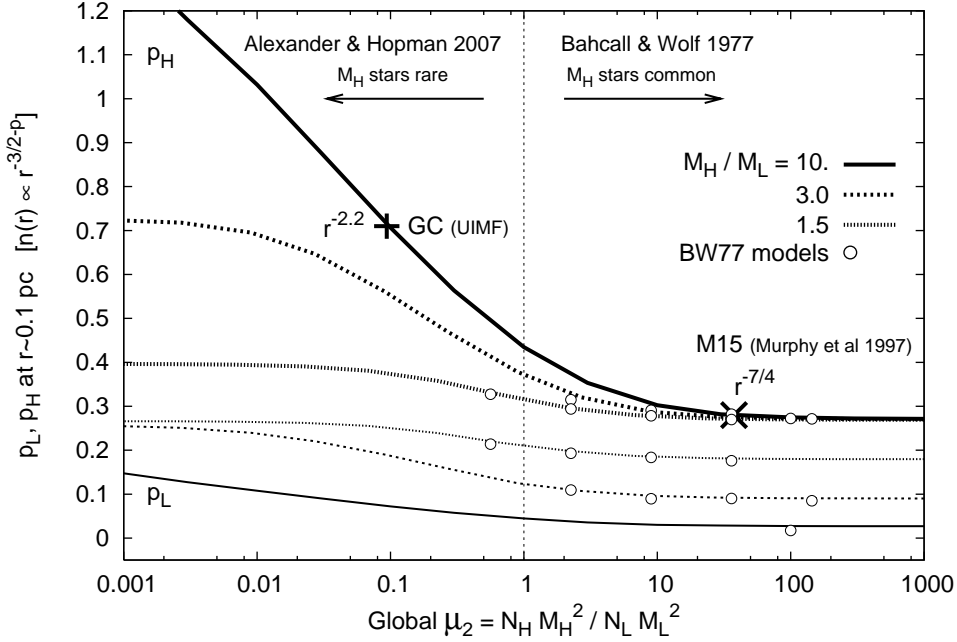}\tabularnewline
\end{tabular}
\par\end{centering}

\caption{\label{f:p12} Fokker-Planck mass-segregation results. The logarithmic
slopes $p_{H}$ and $p_{L}$ of the DFs of the heavy stars (thick
lines) and light stars (narrow lines) , evaluated at ($r\!\sim\!0.1$
pc in the GC), as function of the global relaxational self-coupling
parameter $\mu_{2}$, for mass ratios of $M_{H}/M_{L}\!=\!1.5,\,3,\,10$
(Alexander \& Hopman 2007, in prep.). The logarithmic slope of the
stellar density of massive stars in the GC, assuming a universal IMF
and continuous star formation history ($\mu_{2}\!\sim\!0.13$, Fig.
\ref{f:mu2}, $\alpha_{H}\!=\!3/2+p_{H}\!\simeq\!2.2$) is indicated
by a cross on the left, and for globular cluster M15 (assuming it
harbors an IMBH), on the right(estimated at $\mu_{2}\!\sim\!37$,
based on the mass function model of \citet{mur+97}, $\alpha_{H}\!\simeq\!1.75$).
The results for the models calculated by \citet{bah+77} are indicated
by circles. }

\end{figure}

\textcolor{black}{\begin{equation}
\frac{\partial}{\partial\tau}g_{M}(x,\tau)=-x^{5/2}\frac{\partial}{\partial x}Q_{M}(x,\tau)-R_{M}(x,\tau)\,,\label{e:FPeq}\end{equation}
where $M$, $x$ and $\tau$ are} the dimensionless mass, energy,
and time, respectively, $g_{M}$ is the dimensionless DF, $Q_{M}$
is the flow integral, which expresses the diffusion rate of stars
by 2-body scattering to energies above $x$, and $R_{M}\!\propto\! g_{M}/T_{R}$
is the $J$-averaged effective loss-cone term. $Q_{M}$ and $R_{M}$
are non-linear functions of the set of DFs $\left\{ g_{M}\right\} $.
The equations are solved for $\left\{ g_{M}\right\} $ by finite difference
methods starting from an arbitrary initial DF and integrating forward
in time until steady state is reached, subject to the boundary conditions
that no stars exist at energies above some destruction energy $x_{D}$,
$g_{M}(x\!>\! x_{D})\!=\!0$, and that the unbound stars are drawn
from an isothermal distribution with a given mass function, $g_{M}(x\!<\!0)\!=\! N_{M}\exp$($Mx)$.
\citet{bah+77} showed that the stellar space density distribution,
\begin{equation}
n_{M}(r)\!\propto\!\int_{-\infty}^{r/r_{h}}g_{M}(x)\sqrt{r/r_{h}-x}\mathrm{d}x\,,\label{e:nm}\end{equation}
does not depend strongly on the exact form of the loss-cone term,
and proceeded to use in their mass-segregation calculations a simplified
version of Eq. (\ref{e:FPeq}) by setting $R_{M}\!=\!0$. This approximation
can be justified by noting that while the existence of a loss-cone
drastically increases the flow rate of stars into the MBH, it typically
affects only a small volume in phase space near $J\sim0$. This translates
to small changes only in the $J$-integrated DF $g_{M}(x)$, mostly
for $x\rightarrow x_{D}$, and even smaller changes in $n_{M}(r)$
due to the smoothing effect of the $g_{M}(x)\!\rightarrow\! n_{M}(r)$
transformation (Eq. \ref{e:nm}). Here we adopt this approximation
to allow direct comparison with the \citet{bah+77} results, after
verifying that the inclusion of the loss-cone term indeed does not
significantly change the derived stellar cusps (cf Fig. \ref{f:GCmseg}
and \ref{f:p12}).

We calculated a suite of such Fokker-Planck mass-segregation models
for 2-mass populations with different mass ratios $M_{H}/M_{L}$ and
mass functions $N_{H}/N_{L}$, spanning a very wide range of the global
relaxational self-coupling parameter values%
\footnote{It is unlikely that real stellar system will have relaxational self-coupling
parameters $\mu_{2}\!\ll\!0.1$. However, the study of such models
is useful for understanding the mathematical properties of the solution. %
}, $10^{-3}\!<\!\mu_{2}\!<\!10^{3}$ (Alexander \& Hopman 2007, in
prep.). The DFs are not exact power-laws, and the logarithmic slopes
$p_{M}(x)\!=\!\mathrm{d\log}g_{M}/\mathrm{d\log}x$ depend somewhat
on energy, especially near the boundaries. However, analysis of the
results is considerably simplified by the fact that the values of
$p_{M}(x)$ vary monotonically with $\mu_{2}$ at all $x$, and so
the order ranking of $p_{M}$ for different models does not depend
on the choice of $x$. Figure (\ref{f:p12}) shows $p_{L}$ and $p_{H}$
at $x\!=\!10$, which corresponds to $r\!\sim\!0.1$ pc in the GC.
This choice samples $g_{M}(x)$ in a representative region, far from
either boundaries at $x\!=\!0$ and $x_{D}\!=\!10^{4}$, and translates
to an observationally relevant region in the GC, which is close enough
to the MBH to be nearly Keplerian, but still contains a large number
of observed stars to allow meaningful statistics (cf Fig. \ref{f:HBRC}). 

Figure (\ref{f:p12}) shows that for $\mu_{2}\!>\!1$, the Fokker-Planck
calculations recover the Bahcall-Wolf solution: $p_{H}\!\simeq\!1/4$
irrespective of the mass ratio, and $p_{L}\!\simeq\!(1/4)(M_{L}/M_{H})$.
However, for $\mu_{2}\!<\!1$ there is a marked qualitative change
in the nature of the solutions, as anticipated by the analysis in
\S \ref{ss:mu2}. The more the light stars dominate the population
(the smaller $\mu_{2}$), the more they approach the single population
solution $p_{L}\!=\!1/4$ (\S \ref{ss:BW77}). The heavy stars settle
to a much steeper cusp with $p_{H}\!>\!1/4$. Figure (\ref{f:p12})
also shows the grid of models explored by \citet{bah+77}, which,
while large, covers only the $\mu_{2}\!>\!1$ range. This explains
why the strong segregation branch of the solutions escaped their notice
(their one model with $\mu_{2}\!\simeq\!0.6$ has a low mass ratio
$M_{H}/M_{L}\!=\!1.5$, where the two solution branches are not very
different). 

The $\mu_{2}\!<\!1$ models explored here follow the $p_{M}\!\propto\!\Ms$
relation noted by \citet{bah+77} for the approximate mass-segregation
solutions without a loss-cone term. Therefore, in those models where
the limit $p_{L}\!\rightarrow\!1/4$ is reached (for $M_{H}/M_{L}\!=\!1.5$,
$3$), the heavy stars reach the asymptotic value $p_{H}\!\rightarrow\!(1/4)(M_{H}/M_{L})$.
It remains to be seen whether this result also holds for higher mass
ratios, and for the full Fokker-Planck equation (Eq. \ref{e:FPeq})
with the loss-cone term. 

A realistic evolved stellar system, such as the GC, is expected to
have a maximal mass ratio of at least $M_{H}/M_{L}\!=\!10$ and $\mu_{2}\!\simeq\!0.15$
(Fig. \ref{f:mu2}). The mass-segregation calculations indicate that
for these parameters the stellar BHs are expected to form an $\alpha_{H}\!\simeq\!2.1$--$2.2$
cusp (Figs. \ref{f:GCmseg}, \ref{f:p12}), significantly steeper
than the $\alpha_{H}\!=\!1.75$ predicted by the Bahcall-Wolf solution
of moderate segregation. It is encouraging that this logarithmic slope
is very close to that found in numerical simulations (Fig. \ref{f:GCmseg})
and that it is broadly consistent with what is needed to explain the
observed trend in the stellar surface density distributions in the
GC in terms of mass segregation (Fig. \ref{f:HBRC}). In other systems
the moderate segregation solution may apply. For example, if the globular
cluster M15 contains an IMBH \citep[e.g. ][]{ger+02}, then a tentative
determination of its present-day mass function \citep{mur+97} suggests
a high relaxational self-coupling parameter, $\mu_{2}\!\sim\!40$
and a relatively shallow $\alpha\!=\!7/4$ cusp of stellar BHs. Full-scale
numeric simulations that are free of the restrictive assumptions of
the analytic approach adopted here (Keplerian potential, fixed boundary
conditions, approximate treatment of the loss-cone and a fixed 2-mass
stellar population) are needed to verify and test these predictions
in more detail. 

Strong segregation will affect the cosmic rates of EMRI events. A
detailed analysis of the anticipated change relative to the various
discrepant published rate estimates depends on their specific assumptions
(e.g. the assumed mass function, slope of the cusp, normalization
of the stellar number density), and is outside the scope of this review.

\section{Resonant relaxation}

\label{s:RR}

\subsection{Resonant relaxation dynamics}

\label{ss:RR_dyn}

The effect of 2-body relaxation on a test star is incoherent: the
star experiences randomly oriented, uncorrelated perturbations from
the ambient stars, and as a result its orbit deviates in a random-walk
fashion from its original phase-space coordinates (in a stationary
spherical smoothed potential where $E$ and $J$ would have been conserved
in the continuum limit, $\Delta E\!\propto\!\sqrt{t}$ and $\Delta J\!\propto\!\sqrt{t}$
due to 2-body interactions). Resonant relaxation (RR) \citep{rau+96,rau+98}
is a form of accelerated relaxation of the orbital angular momentum,
which occurs when approximate symmetries in the potential restrict
the orbital evolution of the perturbing stars. This happens in the
almost Keplerian potential near a MBH, where the orbits are approximately
fixed ellipses (the potential of the enclosed stellar mass far from
the MBH, or General Relativistic (GR) precession near the MBH, eventually
leads to deviations from pure elliptical orbits), or in a non-Keplerian,
but nearly spherically symmetric potential around a MBH, where the
orbits approximately conserve their angular momentum and move on rosette-like
planar orbits (the fluctuations in the potential due to stellar motions
eventually lead to deviations from strictly planar orbits). As long
as the symmetry is approximately conserved, on times shorter than
the coherence timescale $t_{\omega}$, the orbit of a test star with
semi-major axis $a$ experiences correlated (coherent) perturbations%
\footnote{RR is better described as {}``coherent relaxation''. The term {}``resonant''
refers to the equality of the radial and azimuthal orbital periods
in a Keplerian potential, which results in closed ellipse orbits. %
}, which can be described as a constant residual torque exerted by
the superposed potentials of the $\Ns(<\! a)$ randomly oriented elliptical
{}``mass wires'' (in a Keplerian potential ) or {}``mass annuli''
(in a non-Keplerian spherical potential) that represent the orbitally-averaged
mass distribution of individual perturbing stars. The magnitude of
the residual torque is then $\dot{J}\!\sim\!\Ns^{1/2}(<\! a)G\Ms/a$
and the change in the angular momentum of the test star increases
linearly with time, $\Delta J\!\sim\!\dot{J}t$ (for $t\!<\! t_{w}$).
The orbital energy, on the other hand, remains unchanged, since the
potential is constant.

\begin{figure}
\noindent \begin{centering}
\includegraphics[width=0.85\columnwidth]{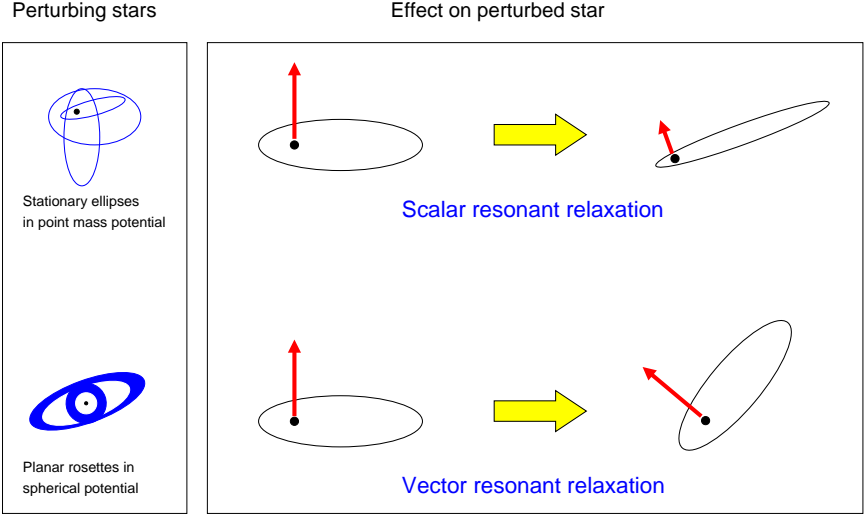}
\par\end{centering}

\caption{\label{f:RRsv}A sketch comparing the symmetries leading to scalar
and vector RR. Top: The torques by fixed elliptical {}``mass wires''
in a Keplerian potential lead to rapid changes in both the direction
and magnitude of the orbital angular momentum of a test star. Bottom:
The torques by fixed {}``mass annuli'' in a non-Keplerian spherical
potential lead to rapid changes in the direction, but not in the magnitude
of the orbital angular momentum of a test star.}

\end{figure}

RR in a Keplerian potential is called \emph{scalar} RR since it changes
both the magnitude and direction of $\mathbf{J}$. Scalar RR can therefore
change a circular orbit into an almost radial, MBH-approaching one.
In contrast, RR in a non-Keplerian spherical potential is called \emph{vector}
RR since, for reasons of symmetry, it changes only the direction of
$\mathbf{J}$, but not its magnitude (Fig. \ref{f:RRsv}). Vector
RR can randomize the orbital orientations, but does not play a role
in supplying stars to the loss-cone.

\begin{figure}
\noindent \begin{centering}
\includegraphics[width=0.85\columnwidth]{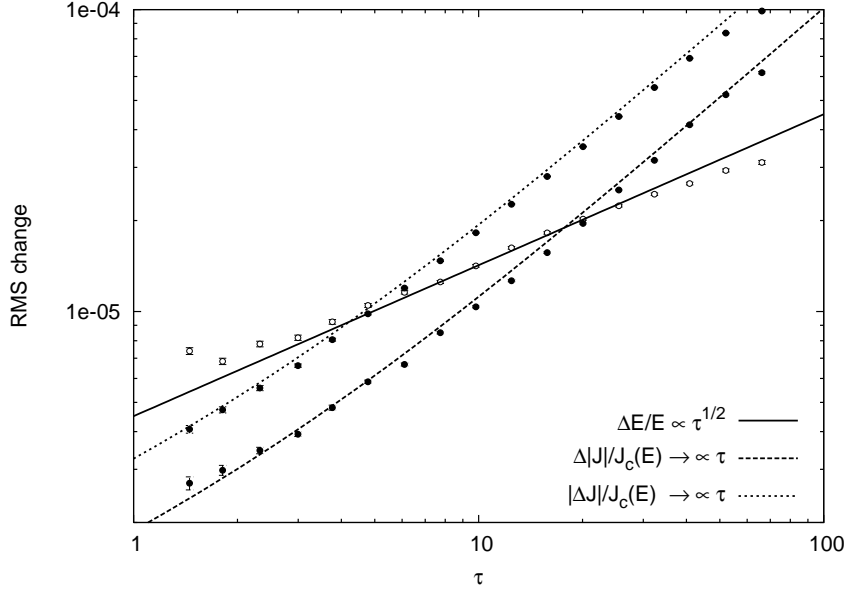}
\par\end{centering}

\caption{\label{f:RRsim} A correlation analysis of $N$-body simulations ($N\!=\!200$)
showing the relaxation of energy and of scalar and vector angular momentum
around a MBH in the Keplerian limit ($\Ms/\Mbh\!=\!3\!\times\!10^{-7}$)
for a thermal population of stars, as function of the elapsed time-lag
(Eilon, Kupi \& Alexander, 2007, in prep.). The change in $\Delta E/E$,
$\Delta J/J_{c}$ and $\left|\Delta\mathbf{J}\right|/J_{c}$ is plotted
as function of the normalized time-lag $\tau\!=\!\Delta t/P$ in the
range $1\!\le\tau\!\le\!100$. The mass precession coherence time
of the system is $\tau_{M}\!=\! t_{M}/P\!\sim\!1.7\times10^{4}$,
and the potential fluctuation coherence time is $\tau_{\phi}\!=\! t_{\phi}/P\!\sim1.2\times10^{5}$,
so both scalar and vector RR are expected to grow linearly over the
plotted range. More detailed analysis shows that the $\Delta J/J_{c}(E)$
is a function of both energy and angular momentum, which for $\tau\!\rightarrow\!0$
scales as $\sqrt{\tau}$, and for $1\!\ll\!\tau\!\ll\!\tau_{w}$ scales
as $\tau$, and that $\left|\Delta\mathbf{J}\right|/J_{c}$ is simply
proportional to $\Delta J/J_{c}$. The correlation analysis is an
efficient method for quantifying relaxation in $N$-body results (cf
\citealt{rau+96}, Fig. 1). The theoretical predictions for $\Delta J/J_{c}$
and $\left|\Delta\mathbf{J}\right|/J_{c}$ fit the numeric results
very well. As expected, $\Delta E/E\!\propto\!\sqrt{\tau}$ at all
time-lags. }

\end{figure}

On timescales longer than the coherence time, the orbital orientations
of the perturbing stars drift, and coherence is lost. the maximal
change in angular momentum during the linear coherence time, $\Delta J_{\omega}\sim\dot{J}t_{\omega}$
then becomes the {}``mean free path'' for a random walk in $J$-space,
whose time-step is $t_{\omega}$. On timescales longer than the coherence
time, the angular momentum changes incoherently $\propto\!\sqrt{t}$,
but much faster than it would have in the absence of RR. The energy
is unaffected by RR and always evolves incoherently $\propto\!\sqrt{t}$
on the long non-resonant relaxation timescale (Fig. \ref{f:RRsim}).
The RR timescale $T_{RR}$ is defined, like the incoherent 2-body
relaxation timescale, as the time to change $J$ by order $J_{c}$
(Eqs. \ref{e:TE}, \ref{e:TJ}), $T_{RR}\!\sim\!\left(J_{c}/\Delta J_{\omega}\right)^{2}t_{\omega}$,
which can be expressed as \citep{hop+06a}\begin{equation}
T_{RR}=A_{RR}^{\omega}\frac{\Ns(<a)}{\mu^{2}(<a)}\frac{P^{2}(a)}{t_{\omega}}\simeq\frac{A_{RR}^{\omega}}{\Ns(<a)}\left(\frac{\Mbh}{\Ms}\right)^{2}\frac{P^{2}(a)}{t_{\omega}}\,,\label{e:TRRtw}\end{equation}
where $\mu$ is the relative enclosed enclosed stellar mass, $\mu\!=\!\Ns\Ms/\left(\Mbh+\Ns\Ms\right)$,
$P$ is the radial orbital period, and the last approximate equality
holds in the Keplerian regime. Here and below, the constants $A_{RR}^{\omega}$
are numerical factors of order unity that depend on the specifics
of the coherence-limiting process, on the orbital characteristics
of the test star, and probably also on the parameters of the stellar
distribution. These constants are not well-determined at this time.

The coherence time depends on the symmetry assumed and on the process
that breaks it. For a non-relativistic near-Keplerian potential, the
limiting process is precession due to the potential of the distributed
stellar mass%
\footnote{The enclosed stellar mass $\Ns\Ms$ changes the Keplerian period $P\!\propto\!\Mbh^{-1/2}$
by $\Delta P/P$$=\!\Ns\Ms/2\Mbh$\\
$=\!\Delta\varphi/2\pi$. Identifying de-coherence with a phase drift
$\Delta\varphi\!=\!\pi$ then implies $t_{M}\!\sim\!(\pi/\Delta\varphi)P$.%
}, \begin{equation}
t_{\omega}=t_{M}\sim\frac{\Mbh}{\Ns(<a)\Ms}P(a)\,.\end{equation}
Remarkably, the resulting RR timescale does not depend on $\Ns$.
Close to the MBH is much shorter than the non-coherent 2-body relaxation
timescale (here denoted for emphasis as $T_{NR}$),\begin{equation}
T_{RR}^{M}=A_{RR}^{M}\frac{\Mbh}{\Ms}P(a)\sim\frac{\Ns(<a)\Ms}{\Mbh}T_{NR}\,.\end{equation}
Yet closer to the MBH, it is GR precession that limits the coherence, 

\begin{equation}
t_{\omega}=t_{GR}=\frac{8}{3}\left(\frac{J}{J_{\mathrm{LSO}}}\right)^{2}P(a)\,,\end{equation}
where $J_{\mathrm{LSO}}\!=\!4G\Mbh/c$ is the last stable orbit for
$\epsilon\!\ll\! c^{2}$. The GR precession is prograde, while that
due to the distributed mass is retrograde, and so they may partially
cancel each other. Their combined effect on the scalar RR timescale
is \begin{equation}
T_{RR}^{s}\simeq\frac{A_{RR}^{s}}{\Ns(<a)}\left(\frac{\Mbh}{\Ms}\right)^{2}P^{2}(a)\left|\frac{1}{t_{M}}-\frac{1}{t_{\mathrm{GR}}}\right|\,.\end{equation}
Since $t_{M}$ increases with $r$, while $t_{\mathrm{GR}}$ decreases
with $r$, scalar RR is fastest at some finite distance from the MBH,
which typically coincides with $\sim\! r_{\mathrm{crit}}/2$ for LISA
EMRI targets (Fig. \ref{f:GC_RR}). 

Precession does not affect vector RR. The coherence in a non-Keplerian
spherical potential is limited by the change in the total gravitational
potential $\phi\!=\!\phi_{\bullet}+\phi_{\star}$ caused by the fluctuations
in the stellar potential, $\phi_{\star}$, due to the realignment
of the stars as they rotate by $\pi$ on their orbits, 

\begin{equation}
t_{\omega}=t_{\phi}=\frac{\phi}{\dot{\phi_{\star}}}\sim\frac{\Ns^{1/2}(<a)}{2\mu}P(a)\simeq\frac{\Mbh}{2\Ns^{1/2}(<a)\Ms}P(a)\,,\end{equation}
where the last approximate equality holds in the Keplerian regime.
The vector RR timescale is obtained by substituting $t_{\phi}$ in
Eq. (\ref{e:TRRtw}), \begin{equation}
T_{RR}^{v}=2A_{RR}^{v}\frac{\Ns^{1/2}(<a)}{\mu(<a)}P(a)\simeq2A_{RR}^{v}\left(\frac{\Mbh}{\Ms}\right)\frac{P(a)}{\Ns^{1/2}(<a)}\,.\end{equation}
 Vector RR is much faster than scalar RR (Fig. \ref{f:GC_RR}).

\subsection{Resonant relaxation and EMRI rates}

\label{ss:RR_EMRI}

The efficiency of scalar RR quickly decreases with distance from the
MBH, since the coherence time falls as $\Ms(<r)/\Mbh$ grows. At $r_{h}$,
where $\Ms(<r_{h})/\Mbh\!\sim\!{\cal O}(1)$, scalar RR is almost
completely quenched. Because $r_{\mathrm{crit}}\!\sim\! r_{h}$ for
tidal disruption (\citealt{lig+77}; \S \ref{ss:LC}), RR does not
significantly enhance the tidal disruption rate \citep{rau+96}. In
contrast, $r_{\mathrm{crit}}\!\sim\!0.01$ pc for EMRI events, where
$\Ms(<r_{\mathrm{crit}})/\Mbh\!\ll\!1$ and $T_{RR}^{s}$ is near
its minimum. Scalar RR therefore dominates the dynamics of the loss-cone
for GW EMRI events \citep{hop+06a}. Scalar RR accelerates the flow
of stars in phase-space from large-$J$ orbits to low-$J$ orbits
that approach the MBH and can lose orbital energy and angular momentum
by the emission of GWs. However, if unchecked, RR would continue to
rapidly drive the stars to plunging orbits that fall directly into
the MBH. This is where GR precession is predicted to play an important
role \citep{hop+06a}. Orbits with very small periapse, $r_{p}\!\sim\!\mathrm{few}\!\times\! r_{s}$,
where GW emission becomes appreciable, are also orbits where the GR
precession rate becomes large enough ($t_{\mathrm{GR}}$ becomes short
enough) to quench RR, and allow the EMRI inspiral to proceed undisturbed.
This subtle cancellation, which is critical for the observability
of EMRI events, still has to be verified by direct simulations.

\begin{figure}
\noindent \begin{centering}
\includegraphics[width=0.75\columnwidth]{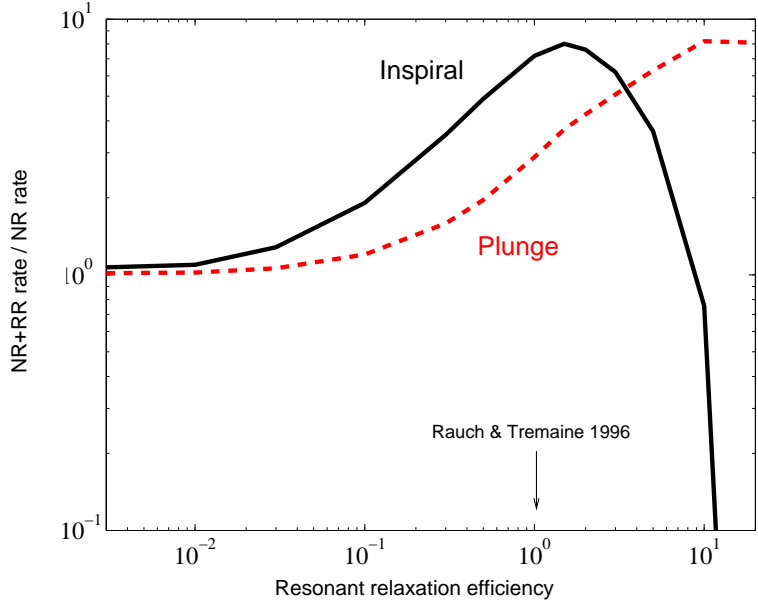}
\par\end{centering}

\caption{\label{f:ChiRR}The relative rates of GW EMRI events and direct infall
(plunge) events, as function of the unknown efficiency of RR, $\chi$,
normalized to $\chi\!=\!1$ for the values derived by \citet{rau+96}.}

\end{figure}

The effect of scalar RR can be included in an approximate way in the
Fokker-Planck equation (Eq. \ref{e:FPeq}) as an additional loss-cone
term $R_{RR}\!\propto\!\chi g/T_{RR}^{s}$, where the efficiency factor
$\chi$ parametrizes the uncertainties that enter through the various
order-unity factors $A_{RR}^{\omega}$ (Eq. \ref{e:TRRtw}). Such
calculations show that the poorly determined value of the efficiency
can strongly affect the predicted EMRI rates (Fig. \ref{f:ChiRR}).
As the efficiency rises, the EMRI rate first increases because stars
are supplied faster to the loss-cone, but when the efficiency continues
to rise the stars are drained so rapidly into the MBH, that the EMRI
rates are strongly suppressed. 

\citet{rau+96} explored the efficiency of RR by a few small-scale
$N$-body simulations, and noted a large variance around the derived
mean efficiency. Here we use their mean efficiency as the reference
point ($\chi\!=\!1$), but consider also values smaller and larger.
Figure (\ref{f:ChiRR}) shows the GW inspiral rate and direct plunge
rate as function of the unknown efficiency $\chi$, relative to the
no-RR case ($\chi\!=\!0$), derived from Fokker-Planck calculations
of a single mass population. The EMRI rate rises to $\sim\!8$ times
more than is expected without RR, peaking at $1\!\lesssim\chi\!\lesssim\!2$,
but then falls rapidly to zero at $\chi\!\gtrsim\!10$. The strong
$\chi$-dependence of the EMRI rates provides strong motivation to
determine the RR efficiency and its dependence on the parameters of
the system both numerically (\citealt{gur+07}; Eilon, Kupi \& Alexander
2007, in prep; Fig \ref{f:RRsim}), and by direct observations of
the only accessible system at present where RR effects may play a
role---the stars around the Galactic MBH. 

\subsection{Resonant relaxation and stellar populations in the GC}

\label{ss:RR_GC}

The stellar population in the GC includes both young and old stars,
and is composed of distinct sub-populations, each with its own kinematical
properties (see \citealt{ale05} for a review). As shown below, RR
can naturally explain some of the systematic differences between the
various dynamical components in the GC. Conversely, GC observations
of these populations can then test the various assumptions and approximations
that enter into analytic treatment of RR, and in particular constrain
the poorly determined RR efficiency.

\begin{figure}
\noindent \begin{centering}
\includegraphics[width=0.85\columnwidth]{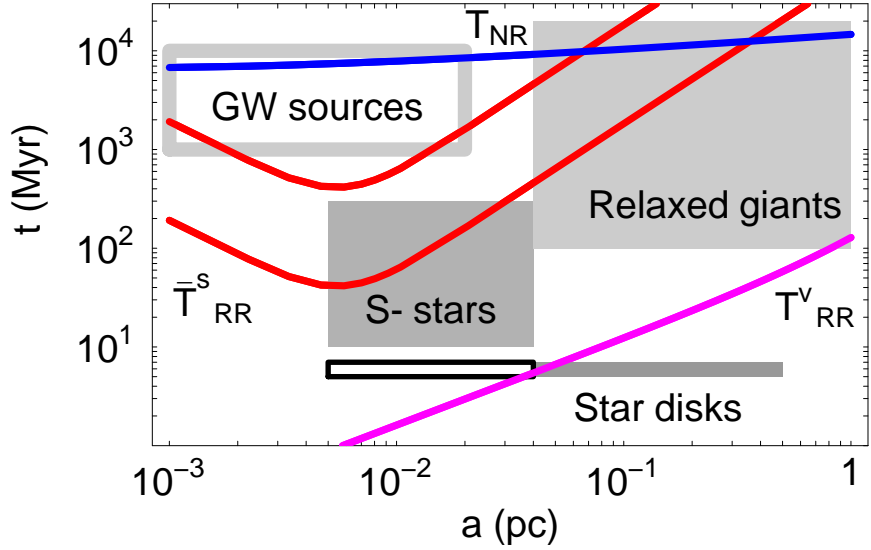}
\par\end{centering}

\caption{\label{f:GC_RR} Evidence for resonant relaxation in the GC in the
age .vs. distance from the MBH plane. The spatial extent and estimated
age of the various dynamical sub-populations in the GC (shaded areas)
is compared with the non-resonant 2-body relaxation timescale (top
line, for assumed mean mass of $\Ms\!=\!1\,\Mo$) and with the scalar
RR timescale (two curved lines, top one for $\chi\Ms\!=\!1\,\Mo$,
bottom one for $\chi\Ms\!=\!10\,\Mo$) and vector RR timescale (bottom
line, for $\chi\Ms\!=\!1\,\Mo$). The populations include the young
stellar rings in the GC (filled rectangle in the bottom right); the
S-stars, if they were born with the disks (open rectangle in the bottom
left); the maximal lifespan of the S-stars (filled rectangle in the
middle left); the dynamically relaxed old red giants (filled rectangle
in the top right); and the reservoir of GW inspiral sources, where
the age is roughly estimated by the progenitor's age or the time to
sink to the center (open rectangle in the top left). Stellar components
that are older than the various relaxation times must be randomized.
(\citealt{hop+06a}, reproduced with permission from the \emph{Astrophysical
Journal}).}

\end{figure}

Figure (\ref{f:GC_RR}) summarizes the typical distance scales and
ages associated with these populations, and compares them with the
various relaxation timescales. The calculation of the relaxation timescales
are approximate since they assume a single mass population. The non-resonant
2-body relaxation timescale (Eq. \ref{e:TE}) is roughly independent
of radius in the GC, $T_{NR}\!\sim\!\mathrm{few}\times10^{9}$ yr
(assuming a mean stellar mass of $\Ms\!=\!1\,\Mo$; in a multi-mass
system it is expected to decrease to $T_{NR}\!\sim\!10^{8}$ yr in
the inner 0.001 pc due to mass segregation, \citealt{hop+06b}). Because
neither the RR efficiency, nor the mass function is known with confidence,
the scalar RR timescale, $T_{RR}^{s}$, is shown for two different
assumptions; $\chi\Ms\!=\!1\,\Mo$ and $\chi\Ms\!=\!10\,\Mo$. As
discussed in \S  \ref{ss:RR_dyn}, beyond $r\!\sim\!0.1$ pc, $T_{RR}^{s}\!>\! T_{NR}$
due to mass precession, and the loss-cone replenishment is dominated
by non-coherent relaxation. $T_{RR}^{s}$ decreases toward the MBH,
until it reaches a minimum, where it starts increasing again due to
GR precession. The distance scale where $T_{RR}^{s}$ is shortest
happens to coincide with the volume $r\!\lesssim\! r_{\mathrm{crit}}$,
where most GW EMRI sources are expected to lie and where $T_{RR}^{s}\!\ll\! T_{NR}$,
so RR dominates EMRI loss-cone dynamics (\S \ref{ss:RR_EMRI}). In
contrast to scalar RR, the vector RR timescale $T_{RR}^{v}$ (shown
here for an assumed $\chi\Ms\!=\!1\,\Mo$) decreases unquenched toward
the MBH. 

Dynamical populations and structures whose estimated age exceed these
relaxation timescales must be relaxed. Those whose age cannot be determined,
but whose lifespan exceeds the relaxation timescales may be affected,
unless we are observing them at an atypical time soon after they were
created. The youngest dynamical structure observed in the GC is the
stellar disk (or possibly two non-aligned disks) \citep{lev+03,gen+03a,pau+06},
which is composed of $\sim\!50$ young massive OB stars with an age
of $t_{\star}\sim\!6\!\pm\!2$ Myr, on co-planar, co-rotating orbits
that extend between $\sim\!0.04$--$0.5$ pc. The inner edge of the
disk is sharply defined and it coincides with the outer boundary of
the S-stars cluster (\S \ref{ss:MP_GC}). Figure (\ref{f:GC_RR})
shows that the vector RR timescale equals the age of the stellar disk
at its inner edge, and so is consistent with the spatial extent of
the disk. Even if the S-stars were initially the inner part of the
disk (this does not appear likely given that they are systematically
lighter than the disk stars), vector RR would have efficiently randomized
their orbital inclination. However, their measured high eccentricities
\citep{eis+05,ghe+05} would then be hard to explain. If instead,
the S-stars are not-related to the disks, but were tidally captured
around the MBH by 3-body exchange interactions (\S \ref{ss:MP_imp}),
then only their lifespan can be determined. Tidal capture leads to
an extremely eccentric captured orbit (Eq. \ref{e:acapture}). Scalar
RR could then randomize and decrease the eccentricities of at least
a few of the older S-stars closer to the MBH. Vector and scalar RR
could also explain why the old evolved giants (with progenitor masses
of $\Ms\!\sim\!2$--$8\,\Mo$, \citealt{gen+94}) at $r\!\gtrsim\!\mathrm{0.1}$
pc appear dynamically relaxed \citep{gen+00}, in spite of the fact
that their lifespans are shorter than the non-coherent relaxation
time.

It should be noted that the effect of RR on the stellar density distribution
is not expected to be large even quite close to the MBH ($r\!\lesssim\!0.1$
pc), unless the efficiency $\chi$ is very high, because the RR-induced
depletion of the DF at high energies is smoothed by the transformation
from the DF to $n_{\star}(r)$ (Eq. \ref{e:nm}) and by the contribution
of unbound stars to the central density.

\section{Summary}

\label{s:summary}

Relaxation processes play an important role in the GC, where the 2-body
relaxation time is shorter than the age of the system and the stellar
density is high. The scaling laws that follow from the $\Mbh/\sigma$
relation imply that the same must hold for all galaxies with $\Mbh\!\lesssim\!\mathrm{few}\times10^{7}\,\Mo$.
Relaxation processes affect the distribution of stars and compact
remnants, lead to close interactions between them and the MBH, and
may be related to the unusual stellar populations that are observed
in the GC. These are of relevance because of the very high quality
stellar data coming from the GC, and because galactic nuclei with
low-mass MBHs like the GC are expected to be important GW EMRI targets
for the next generation of space borne GW detectors. In addition,
efficient relaxation mechanisms that operate and can be studied in
the GC may play a role even in galactic nuclei with high-mass MBHs,
where 2-body relaxation is unimportant.

Three processes beyond minimal two-body relaxation were discussed
here: accelerated loss-cone replenishment by MPs, strong mass-segregation
in evolved populations, and rapid RR. Evidence was presented that
these processes operate and may even dominate relaxation and its consequences
in the GC: The S-stars and HVSs are consistent with relaxation by
GMCs; there are hints for strong mass segregation in the central density
suppression of the low-mass Red Clump giants and in numeric simulations
of the GC, and RR appears to play a role in the truncation of the
stellar disks and the orbital randomization of the S-stars and the
late type giants. There are also cosmic implications: MPs enable the
efficient merger of binary MBHs, and boost the rates of white dwarf
EMRIs captured near the MBH by tidal disruptions of stellar binaries.
Strong segregation, and in particular RR can strongly affect the EMRI
rates from stellar BHs. 

The stellar dynamics laboratory in the GC holds great promise for
future progress in understanding these mechanisms and their implications. 

\bibliographystyle{cupconf}

\end{document}